\documentclass[11pt,draftcls,onecolumn,journal]{IEEEtran}
\usepackage{graphicx}
\usepackage{epsfig}
\usepackage{latexsym}
\usepackage{amsfonts}
\usepackage{here}
\usepackage{rawfonts}
\usepackage[latin1]{inputenc}
\usepackage[T1]{fontenc}
\usepackage{calc}
\usepackage{url}
\usepackage{enumerate}
\usepackage{color}
\usepackage[tbtags]{amsmath}
\usepackage{amssymb}
\usepackage{upref}
\usepackage{epic,eepic}
\usepackage{times}
\usepackage{dsfont}
\usepackage{comment}
\usepackage{cite}
\usepackage{subcaption}
\usepackage{graphicx}
\usepackage[font={small}]{caption}
\usepackage[font={small}]{subcaption}
\usepackage{amssymb}
\usepackage{bm}
\usepackage{adjustbox}
\usepackage{pifont}
\usepackage{tikz}
\usepackage{multirow}
\usepackage{booktabs}
\usepackage{tikz}
\usepackage{tcolorbox}
\usepackage{stfloats}
\usepackage{mathtools}
\usepackage{amsmath}
\usepackage{adjustbox}
\usepackage{algorithm}
\usepackage{algpseudocode}
\usepackage{tabularx}
\newcolumntype{Y}{>{\raggedright\arraybackslash}X}
\usepackage{tikz}
\usetikzlibrary{positioning, arrows.meta, shapes}

\usetikzlibrary{positioning, arrows.meta}
\definecolor{violetblue}{rgb}{0.3,0.1,0.7}
\definecolor{darkgreen}{rgb}{0.0, 0.5, 0.0} 
\begin{document}
	
	\title{Over-the-Air Federated Learning: Rethinking Edge AI Through Signal Processing}
	
	\author{{Seyed Mohammad Azimi-Abarghouyi,~\IEEEmembership{Member,~IEEE}, Carlo Fischione,~\IEEEmembership{Fellow,~IEEE}, and Kaibin Huang, \textit{Fellow, IEEE}}
	\thanks{S. M. Azimi-Abarghouyi and C. Fischione are with the School of Electrical Engineering and Computer Science, KTH Royal Institute of Technology, Stockholm, Sweden (Emails: $\bigl\{$seyaa,\ carlofi$\bigr\}$@kth.se). K. Huang is with the Department of Electrical and Electronic Engineering, The University of Hong Kong, Hong Kong SAR, China (Email: huangkb@hku.hk). {This project has received funding from
		the European Union's Horizon Europe programme under grant agreement
		No. 101137954. The authors would like to thank the rest of the BATTwin
		consortium for supporting this research.}}
}

\markboth{IEEE Signal Processing Magazine,~Vol.~XX, No.~XX, May~2026}%
{\MakeLowercase{\textit{et al.}}: Author Guidelines for Special Issue Articles of IEEE SPM}

\maketitle
\vspace{-5pt}
\section*{Abstract}
\label{sec:abstract}
Over-the-Air Federated Learning (AirFL) is an emerging paradigm that tightly integrates wireless signal processing and distributed machine learning to enable scalable AI at the network edge. 
{By exploiting wireless superposition over a shared multiple-access channel, AirFL turns simultaneous transmissions of local model updates into an analog aggregate at the receiver, thereby reducing communication latency, bandwidth usage, and energy consumption in wireless aggregation domains. This article develops a design-oriented tutorial view of analog AirFL. We organize existing schemes according to the signal-processing mechanism used to enable AirFL aggregation: transmitter-side channel compensation and power control, receiver-side equalization and high-dimensional processing, or learning-aware aggregation weighting. This viewpoint leads to three representative classes---CSIT-aware, blind, and weighted AirFL---and clarifies their assumptions, performance tradeoffs, complexity, and deployment limitations. We further discuss synchronization, digital and hybrid analog-digital realization, and open research directions for integrating AirFL into practical wireless edge-AI systems.}

\section{Introduction}
The convergence of edge AI and wireless communications is shaping the next generation of intelligent, networked systems. At the core of this paradigm lies Federated Learning (FL), a distributed machine learning framework in which devices such as smartphones, IoT sensors, and autonomous vehicles collaboratively train a global model without sharing their raw data~\cite{mcmahan}. The fundamental process involves each device performing local model updates based on its private data and then transmitting these updates to a central server, where aggregation takes place to refine the global model. This cycle is repeated across multiple rounds until convergence.

Originally introduced to preserve privacy and data locality in decentralized environments, FL has rapidly evolved to address broader challenges inherent to large-scale networked systems. However, its real-world deployment remains hindered by a critical limitation: communication overhead. In traditional FL setups, each device must reliably send high-dimensional model updates to the central server at every round. These frequent transmissions often occur over limited-bandwidth, energy-constrained wireless links, resulting in a significant mismatch between the communication-intensive nature of FL and the capacity of existing wireless networks---especially as the number of participating devices grows.

To overcome this bottleneck, Over-the-Air Computation (AirComp) has emerged as a transformative solution~\cite{nazer, gold}. Unlike conventional methods that recover individual messages, AirComp exploits the superposition property of wireless signals to compute desired functions---such as sums---directly over the Multiple Access Channel (MAC). AirComp should thus be viewed as one component within the broader problem of FL, where its function-aggregation capability is leveraged as a building block toward a larger objective. This forms the foundation of Over-the-Air Federated Learning (AirFL), a new class of FL systems where all devices transmit simultaneously, and the server receives a naturally aggregated signal. By collapsing communication and aggregation into a single step, AirFL significantly reduces latency, bandwidth consumption, and energy usage. This makes it a highly promising approach for achieving scalable, efficient FL in resource-constrained wireless environments. However, realizing this potential requires machine-learning-aware wireless signal-processing techniques---capable of handling statistical and computational heterogeneity across devices and of providing unbiased (or controllably biased) aggregation while ensuring privacy, fairness, and convergence.

{This gain, however, comes from treating the wireless channel as a computing medium rather than merely as a transport pipe, and this changes the deployment assumptions relative to conventional digitally routed FL. In standard FL, devices may be geographically dispersed as long as their decoded updates can be delivered to the server through the communication network. In AirFL, by contrast, aggregation occurs directly in the received waveform, without decoding individual device updates. Therefore, the devices participating in one aggregation round must be coordinated through a common wireless aggregation interface, such as a cell, an edge cluster, an IoT gateway region, a factory floor, a vehicular platoon, or a coordinated antenna area. They need not be physically adjacent or experience identical channels: channel heterogeneity in gain and phase can be partially compensated by transmitter-side channel inversion, receiver equalization, power control, or spatial processing. Nevertheless, the devices must share a common aggregation opportunity, i.e., a time-frequency resource observed by a receiver, or a coordinated set of receivers, with an appropriate synchronization level for the intended aggregation mechanism. As a result, direct AirFL is most natural in localized or tightly coordinated deployments, whereas geographically dispersed FL requires additional architectural mechanisms such as hierarchical aggregation, relay-assisted aggregation, cell-free reception, or multi-server coordination \cite{netmag}. This also explains why AirFL is not a simple software-layer addition to current wireless systems: most deployed protocols are digital and packet based, and are designed to separate devices and decode individual packets rather than preserve and exploit their waveform superposition.}

{Once this architectural boundary is recognized, the central signal-processing problem becomes clear: the received superposition must represent the desired model aggregate, not an arbitrary channel-distorted mixture of local updates.} Unlike orthogonal schemes, AirFL is highly sensitive to channel impairments, synchronization errors, and hardware limitations. In particular, the accuracy of signal aggregation depends on the alignment of transmitted signals. One early approach to address this relies on accurate Channel State Information at the Transmitter (CSIT) and power control to compensate for channel gains and phase shifts. However, acquiring and maintaining CSIT in practical wireless environments---especially in large-scale or rapidly varying systems---remains a major challenge. These difficulties have sparked growing interest in methods that require only local CSIT~\cite{localCSIT}. Nonetheless, approaches based on global CSIT~\cite{globalCSIT} continue to be explored due to their potential for achieving optimal performance.

More recently, CSIT-free or CSIT-light approaches---most notably blind AirFL~\cite{blind} and weighted AirFL~\cite{wafel}---have emerged as promising practical alternatives. These approaches avoid CSIT-based power control by relying instead on constant-power transmission, limited phase information when needed, massive MIMO and high-dimensional processing at the server, or adaptive weighting within aggregation. Their main practical appeal is reduced edge-device burden and lower coordination overhead, at the cost of stronger receiver processing, statistical assumptions, partial phase stability, or possible aggregation bias.

{These observations motivate the organization of this article. Rather than classifying AirFL schemes only by the amount of channel knowledge they require, we organize them according to where the main compensation for wireless distortion is performed: at the transmitters through power control, at the receiver through equalization and high-dimensional processing, or in the learning layer through aggregation-weight design. This leads to the three representative classes studied in this tutorial: CSIT-aware, blind, and weighted AirFL.}

{Two implementation aspects are then treated across this taxonomy rather than as separate AirFL classes. Synchronization is a prerequisite for analog aggregation, since timing, frequency, and phase offsets determine whether the received waveform can be interpreted as a useful sum. Digital and hybrid analog-digital realization addresses a different but equally important question: how AirFL principles can coexist with coded modulation, packet framing, scheduling, and reliability mechanisms in practical wireless systems. This separation keeps the main taxonomy centered on aggregation design while still accounting for the mechanisms needed to implement AirFL in real networks.}

\subsection{Related Work and Contributions of this Tutorial}

{Previous tutorials and surveys have mainly followed two separate directions. Some focus on FL, emphasizing distributed optimization, privacy, data heterogeneity, and system-level learning architectures~\cite{smith, israel}. Others focus on AirComp, developing the communication-theoretic and signal-processing foundations of function computation over MACs~\cite{alphan, saeedmag}. These works provide important background, but they leave open the need for a unified AirFL treatment that connects AirComp mechanisms with FL convergence and design at the communication--learning interface. This gap calls for a fundamental rethinking of how learning-oriented objectives, metrics, and influencing factors should be integrated into wireless aggregation design.}

{While there are a few brief overview and perspective papers on AirFL~\cite{over1, over2}, as well as broader surveys on distributed and federated learning over wireless networks~\cite{over3, over4, tao_feel6g}, their emphasis is mainly on mapping the field and discussing AirFL at a high level. In particular, the recent 6G-oriented survey in~\cite{tao_feel6g} positions AirFL within a broad federated-edge-learning landscape, together with discussions on task-oriented communication, resource management, synchronization, digital AirComp, RIS-assisted aggregation, UAV-assisted aggregation, and future 6G directions. These works are valuable for survey-level positioning, but they do not provide a unified mathematical and design-oriented treatment of AirFL schemes.}

{The main distinction of this article is that it develops AirFL itself under a common system model and analytical framework. We introduce a taxonomy and present representative CSIT-aware, blind, and weighted AirFL schemes through aggregation models, aggregation-error decompositions, MSE expressions, convergence bounds, and the resulting design criteria. This structure gives a lecture-note-style view of how different AirFL mechanisms operate, what assumptions they require, how their performance tradeoffs arise, and where their complexity and deployment limitations appear.}

{A central theme is convergence-guided AirFL design. Convergence analysis is used not only to assess the effect of wireless aggregation errors, but also to motivate device selection, receiver equalization, aggregation-weight optimization, and the treatment of computational heterogeneity. This viewpoint connects the learning objective with the underlying signal-processing design, rather than treating communication errors only as external perturbations.}

{The article proceeds as follows. We first review AirComp fundamentals and introduce the FL aggregation setup used throughout the tutorial. We then discuss synchronization as a cross-cutting condition for wireless model aggregation. The main body develops CSIT-aware, blind, and weighted AirFL under a common analytical framework, followed by a discussion of digital and hybrid implementations, experimental comparison, and open research directions.}

\section{Over-the-Air Computation Fundamentals}
This section presents the fundamental principles of AirComp over the standard MAC.

\subsection{Signal Model over the MAC}

Consider a wireless system comprising $K$ single-antenna transmitting devices and a receiver with $M$ antennas. This is the most widely adopted wireless model, as devices typically have limited hardware capabilities and physical constraints that restrict them to a single antenna. Each transmitter $k$ sends a scaled version of its data symbol $x_k$ using a complex precoding scalar
\begin{align}
	p_k = |p_k| e^{j\angle p_k},
\end{align}
which controls both the transmission power and phase. The choice of $p_k$, referred to as \textit{power control}, must satisfy either the average power constraint $\mathbb{E}[|p_k|^2] \leq P$ or the instantaneous constraint $|p_k|^2 \leq P$. The received signal vector $\mathbf{y} \in \mathbb{C}^M$ at the receiver is then given by
\begin{align}
	\label{MACsig}
	\mathbf{y} = \sum_{k=1}^{K} \mathbf{h}_{k} p_k x_k + \mathbf{z}.
\end{align}
Here, $\mathbf{h}_k = [h_{k,1}, \ldots, h_{k,M}]^T$ denotes the channel vector from device $k$ to the receiver, with each element
\begin{align}
	h_{k,m} = |h_{k,m}| e^{j\angle h_{k,m}},
\end{align}
representing the channel coefficient from device $k$ to antenna $m$, comprising channel gain $|h_{k,m}|$ and phase $\angle h_{k,m}$. The channels are assumed to remain constant during a given transmission but may vary across different transmissions. The vector $\mathbf{z} \sim \mathcal{CN}(0, \sigma_\text{z}^2 \mathbf{I})$ models complex additive white Gaussian noise (AWGN) at the receiver. On the other hand, to estimate a desired function of the transmitted data, the receiver applies an equalization vector $\mathbf{b} \in \mathbb{C}^M$, yielding the scalar output
\begin{align}
	f = \mathbf{b}^H \mathbf{y} = \mathbf{b}^H \left( \sum_{k=1}^{K} \mathbf{h}_k p_k x_k + \mathbf{z} \right).
\end{align}

\subsection{Nomographic Function Computation}

More generally, AirComp aims to compute a function of the form
\begin{align}
	\label{aircompfunction}
	f = \Phi \left( \sum_{k=1}^{K} \mathbf{h}_k \Psi_k(x_k) + \mathbf{z} \right),
\end{align}
where $\Psi_k(\cdot)$ is the pre-processing function applied at device $k$ and $\Phi(\cdot)$ is a post-processing function applied at the receiver. In AirComp, $\Psi_k$ determines how each device maps its data to the transmitted waveform before superposition, while $\Phi$ determines how the receiver extracts the desired function from the received waveform. The objective of AirComp is to make~\eqref{aircompfunction} as close as possible to a nomographic target function $F$ of the form
\begin{align}
	\label{numographic}
	F = \Theta \left( \sum_{k=1}^{K} \Lambda_k(x_k) \right),
\end{align}
where $\Lambda_k(\cdot)$ and $\Theta(\cdot)$ are known functions. Functions such as~\eqref{numographic} are highly expressive and appears in many practical applications across signal processing, distributed computation, and machine learning. Examples include:
\begin{itemize}
	\item Weighted Sum: $\sum_{k=1}^{K} \alpha_k x_k$
	\item Majority Vote: $\text{sign}\left(\sum_{k=1}^{K} \text{sign}(x_k)\right)$
	\item Polynomial: $\sum_{k=1}^{K} a_k x_k^{k-1}$
	\item P-norm: $\left(\sum_{k=1}^{K} |x_k|^p\right)^{1/p}$
\end{itemize}

\subsection{Performance Metric and Optimization Objective}

A common metric to evaluate the accuracy of AirComp is the mean squared error (MSE) between the received function $f$ and the desired function value $F$ as
\begin{align}
	\label{main_mse}
	\text{MSE} = \mathbb{E} \left[|f - F|^2 \right] = \mathbb{E} \left[ \left| \Phi \left( \sum_{k=1}^{K} \mathbf{h}_k \Psi_k(x_k) + \mathbf{z} \right) - \Theta \left( \sum_{k=1}^{K} \Lambda_k(x_k) \right) \right|^2 \right].
\end{align}
{The goal is to design the device-side pre-processing functions $\Psi_k$ and the receiver-side post-processing function $\Phi$ to minimize this MSE. The design of these functions depends on how channel state information (CSI), i.e., knowledge of $\mathbf{h}_k$ for all $k$ in~\eqref{main_mse}, is acquired, where it is available, and where it is used.}

{In uplink systems, CSI is often first obtained at the receiver from pilot or training signals. When this information is used only at the receiver, it is referred to as channel state information at the receiver (CSIR). In this case, devices do not need channel-dependent transmission parameters before sending their data. After the pilots and data symbols are received, the receiver can use the available CSIR to process the superposed signal directly. Thus, CSIR-based designs avoid a transmitter-side feedback loop and place the main signal-processing burden at the receiver.}

{CSIT, in contrast, means that channel knowledge is used to choose the device-side pre-processing before data transmission. In local-CSIT designs, each device knows only its own effective channel to the receiver and adapts its own transmission independently. This local information may be obtained through calibrated TDD reciprocity from a downlink pilot or through limited feedback of only the corresponding device-specific channel information. Local CSIT therefore reduces coordination and feedback overhead, but it generally cannot realize a globally optimized design because each device acts without knowledge of the other devices' channels.}

{In global-CSIT designs, the edge server has system-wide CSI, usually first acquired as CSIR from uplink pilots. The term global CSIT does not mean that every device knows all channel coefficients. Rather, it means that the globally acquired CSI is used to design transmitter-side pre-processing parameters. The server solves a centralized design problem using the channels of all participating devices and then feeds back only the required device-specific transmission parameter to each device. Compared with purely CSIR-based designs, this introduces an additional estimation--optimization--feedback stage before data transmission. This stage can be costly and delay-sensitive, and the resulting parameters may become outdated when the channels vary quickly.}

{These distinctions lead to different AirComp design categories. CSIR-based schemes use CSI only for receiver-side processing. Local-CSIT schemes use device-specific channel knowledge for decentralized transmitter-side pre-processing. Global-CSIT schemes use system-wide CSI for centralized transmitter-side design followed by device-specific feedback. In most AirComp and AirFL studies, $\Psi_k$ and $\Phi$ are chosen in linear forms: the device-side operation becomes power control, while the receiver-side operation becomes equalization, as discussed in Subsection II-A~\cite{blind,wafel,localCSIT,globalCSIT,khabin,khabin2}.}

\subsection{Challenges and Practical Considerations}

AirComp presents several practical challenges:
\begin{itemize}
	\item {The validity of the baseband MAC model in~\eqref{MACsig} depends on whether the transmitted waveforms overlap over the same time--frequency resource and whether residual timing, frequency, and phase offsets can be represented through stable effective channel coefficients. These synchronization conditions and their AirFL-specific implications are discussed in Section~\ref{sec:synchronization_airfl}.}
	
	\item Perfect computation is generally unachievable due to noise and arbitrary channel coefficients, which lead to unavoidable residual errors.
	
	\item {It entails hardware and protocol constraints. Classical AirComp and AirFL are naturally analog and non-orthogonal: devices transmit simultaneously over the same resource block, and the receiver exploits the resulting superposition rather than decoding individual packets. This differs from current digital wireless protocol stacks, which are designed around channel coding, packet detection, link adaptation, retransmission, and individual-message recovery. Digital and hybrid analog-digital implementations are therefore important for practical deployment and are discussed in Section~\ref{sec:digital_aircomp}.}
	
	\item It is inherently restricted to computing certain classes of functions, with prior research primarily concentrating on summation.
\end{itemize}

Despite these limitations, AirComp offers significant benefits:
\begin{itemize}
	\item It substantially reduces communication overhead by computing functions directly over the air, avoiding individual-message recovery and thereby lowering bandwidth use, energy consumption, and latency.
	
	\item It scales efficiently with the number of devices, making it suitable for large-scale systems.
	
	\item It opens up the potential for creating a ``wireless distributed computing platform''---a kind of \textit{computer over the air}.
\end{itemize}

\section{Federated Learning Fundamentals}
{Having introduced AirComp as a physical-layer mechanism for computing functions over MAC, we now turn to the learning problem that AirFL aims to support. This section introduces the FL objective, the local-training notation, and the server-side aggregation rule used throughout the tutorial.}
 
 Consider $K$ devices, where device $k$ possesses its own local (private) dataset ${\cal D}_k$. The learning model is parametrized by the model vector $\mathbf{w} = [w_1,\ldots,w_s]^\top\in \mathbb{R}^{s\times 1}$, where $s$ is the model size. Then, the local loss function of the model vector $\mathbf{w}$ on ${\cal D}_k$ is
\begin{align}
	F_k(\mathbf{w}) =  \frac{1}{D_k}\sum_{\xi_i\in {\cal D}_k}^{}\ell(\mathbf{w},\xi_i),
\end{align}
where $D_k = |{\cal D}_k|$ is the size of the dataset and the function $\ell(\mathbf{w},\xi_i)$ represents the sample-wise loss, measuring the prediction error of $\mathbf{w}$ on the sample $\xi_i$. Following this, the global loss function applied to all distributed datasets, denoted as $\cup_{k=1}^{K} {\cal D}_k$, is 
\begin{align}
	\label{lossfunction}
	F(\mathbf{w}) = \frac{1}{\sum_{k=1}^{K}D_k}\sum_{k=1}^{K} D_k F_k(\mathbf{w}).
\end{align}

The goal of the training procedure is to discover an optimal parameter vector $\mathbf{w}$ that minimizes $F(\mathbf{w})$, expressed as
\begin{align}
	\label{objective}
\mathbf{w}^* = \arg\min_{\mathbf{w}} F(\mathbf{w}).
\end{align}
{A widely used FL framework for solving~\eqref{objective} is FedAvg~\cite{mcmahan}, which serves as the foundational component of many AirFL schemes. In its general form, FedAvg specifies that devices perform local training and that the server periodically averages the resulting local models; it does not mandate a particular local optimizer. Since many AirFL convergence analyses instantiate this framework with local stochastic gradient descent (local SGD)~\cite{stich_local_sgd}, we adopt this common instantiation for notation.} 

Consider a particular round $t \in \left\{0,\ldots,T-1\right\}$, where $T$ denotes the number of rounds. In this round, each device $k$ updates its own learning model through $\tau$ local SGD steps, each based on a randomly sampled mini-batch $\boldsymbol\xi_k^{t,i}$ with size $B$ drawn from ${\cal D}_k$, as
\begin{align}
	\label{update_step}
	\mathbf{w}_{k}^{t,i+1} = {\mathbf{w}_{k}^{t,i}}- \mu \nabla F_k(\mathbf{w}_{k}^{t,i}, \boldsymbol\xi_k^{t,i}), \forall i \in \left\{0,\ldots,\tau-1\right\},
\end{align}
where $\mu$ is the learning rate. Then, each device $k$ uploads the local model $\mathbf{w}_{k}^{t} = {\mathbf{w}_{k}^{t,\tau}}$ to the server for aggregation. For the equal-weight AirFL setting emphasized in this tutorial, the server-side global model is obtained by averaging the local model parameters as
\begin{align}
	\label{agg11}
	\mathbf{w}_{\text{G}}^{t+1} = \frac{1}{K}\sum_{k=1}^{K} \mathbf{w}_{k}^{t}.
\end{align}
Next, the server broadcasts the obtained global model $\mathbf{w}_{\text{G}}^{t+1}$ to the devices, based on which each device $k$ updates its initial state for the next round as $\mathbf{w}_{k}^{t+1,0} = \mathbf{w}_{\text{G}}^{t+1}, \forall k$.

Thus, AirFL applies AirComp to FL, as shown in Fig.~1, where the objective is to compute the following linear nomographic function:
\begin{align}
	\label{simple_agg}
	F = \frac{1}{K} \sum_{k=1}^{K} x_k.
\end{align}
This direct wireless aggregation approach significantly reduces both latency and bandwidth consumption compared to traditional orthogonal transmission schemes, such as TDMA and FDMA, which require a separate resource block for each device. In these schemes, the server must wait to receive individual model updates from all devices before performing the aggregation.

{Before developing the three main AirFL design principles---transmitter power control, receiver equalization, and aggregation-weight design---we first clarify the synchronization assumptions underlying wireless model aggregation.}

\begin{figure}[tb!]
	\vspace{-70pt}
	\hspace{-70pt}\includegraphics[width =8in]{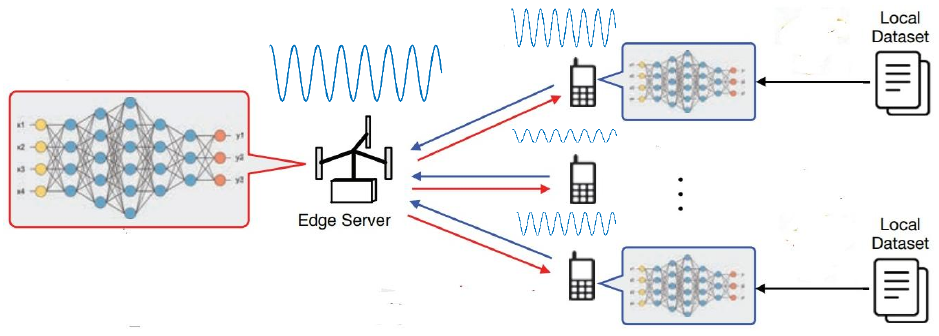}
	\vspace{-620pt} \caption{Standard AirFL setup.}
	\vspace{-10pt}
\end{figure}

\section{{Synchronization Assumptions for AirFL Aggregation}}
\label{sec:synchronization_airfl}

{AirFL aggregation relies on a shared wireless aggregation slot in which the participating devices transmit their local model symbols over the same time--frequency resource. The baseband MAC model in~\eqref{MACsig} assumes that these waveforms overlap at the receiver so that the contribution of device~$k$ to antenna~$m$ can be represented by a complex effective channel coefficient $h_{k,m}$. This effective coefficient may include not only propagation fading, but also residual timing offsets, carrier-frequency offsets, sampling-frequency offsets, phase noise, and hardware calibration errors.}

{This interpretation is useful but should not be misunderstood. Absorbing synchronization errors into effective channel coefficients is an analytical abstraction, not a synchronization solution. It is valid only when the residual offsets remain sufficiently stable over the AirFL aggregation block and when the relevant effective channels can be estimated or statistically handled by the adopted aggregation mechanism. This point is particularly important for phase-sensitive transmitter-side pre-processing: if a transmission coefficient is designed from an effective channel estimated before model transmission, then the same effective channel must remain valid during the model-symbol transmission. Otherwise, the pre-processing can become outdated and may amplify, rather than remove, the aggregation distortion.}

\subsection{{Three Synchronization Requirements}}

{Three forms of synchronization are relevant for AirFL aggregation.}

{\textit{1) Clock synchronization and frame timing:}
	All participating devices must share a common notion of the AirFL communication round and must begin transmission within the intended aggregation slot. If some devices transmit outside this slot, their model symbols do not overlap with those of the other devices, and the received waveform no longer represents the simultaneous aggregation model in~\eqref{MACsig}. This level is analogous to packet/frame-level coordination in conventional wireless systems, but in AirFL it is needed to define the common computation interval.}

{\textit{2) Time alignment of received signals and waveform overlap:}
	The transmitted model symbols must arrive at the server with sufficient symbol-level alignment. This depends on both device clock synchronization and propagation-delay compensation. Small residual timing offsets can be absorbed into the effective channel coefficients when they remain within the guard interval, cyclic prefix, or timing margin of the waveform. However, when the residual delay mismatch becomes comparable to the symbol duration, the server observes inter-symbol interference rather than a coordinate-wise aggregate of the transmitted model entries. In wideband channels, this issue is typically handled using timing advance, cyclic prefixes, guard intervals, OFDM processing, or receive-side equalization.}

{\textit{3) Carrier-frequency and carrier-phase stability:}
	The carrier frequencies and phases of the devices must remain sufficiently stable during the AirFL aggregation block. Residual carrier-frequency offsets, sampling-frequency offsets, and phase noise introduce time-varying rotations into the received signals. If these rotations are approximately constant over the aggregation block, they can be included in the effective channel coefficients. If they vary significantly across the block, different model symbols experience different effective channels, causing aggregation bias, increased MSE, or degraded convergence. This issue is most severe for phase-sensitive designs such as channel inversion and coherent receive equalization.}

\subsection{{Practical Mechanisms for Synchronization and Misalignment Handling}}

{Existing mechanisms for handling synchronization in AirComp and AirFL can be grouped into four categories.}

{First, network-assisted uplink timing control can provide part of the synchronization infrastructure needed by AirComp and AirFL. For example, cellular timing advance adjusts each device's uplink transmission time using propagation-delay estimates obtained from server-side timing references, so that signals from geographically separated users arrive within a prescribed timing window. Existing uplink waveforms such as OFDM, SC-FDMA, or DFT-spread OFDM also provide cyclic prefixes, guard intervals, pilot structures, and receiver-side timing-estimation mechanisms that can absorb residual delay spread and small timing errors~\cite{mahmood_time_sync,localCSIT,guo_waveform,sahin_dfts_ofdm}. These mechanisms do not by themselves create an AirComp protocol, because conventional uplink scheduling typically separates users over orthogonal resources. However, they can provide the timing and waveform-alignment substrate on which AirComp can be implemented by allowing controlled non-orthogonal resource reuse or shared-subchannel transmission among the devices participating in the same aggregation.}

{Second, carrier-frequency and phase control can reduce the time variation of the effective channel during an aggregation block. Relevant mechanisms include oscillator calibration, reference-clock sharing, distributed coherent-transmission protocols, pilot-aided phase tracking, and feedback of residual timing or phase estimates~\cite{airshare,guo_waveform}. These mechanisms are especially important for long model-update transmissions, because even small residual CFO can accumulate a large phase drift across many transmitted symbols.}

{Third, when strict alignment is not feasible, receiver-side misalignment-aware processing can be used. For single-carrier waveforms, matched filtering and sampling can recover useful sums under known or estimated timing offsets. Bayesian AirComp provides another approach by treating the transmitted quantities and misalignment effects statistically and reconstructing the desired aggregate from the received waveform and side information~\cite{shao_misaligned,shao_bayesian}. These methods are relevant when effective synchronization errors cannot be estimated in advance at every edge device.}

{Fourth, digital and noncoherent AirComp can relax strict phase synchronization. Sign-based aggregation, majority-vote AirComp, FSK/PPM-type noncoherent schemes, DFT-spread-OFDM designs, and broadband digital AirComp move part of the burden from coherent analog amplitude/phase alignment to quantization, coding, guard intervals, energy detection, or aggregation-oriented decoding~\cite{gunduz_bpsk,sahin_dfts_ofdm,zhao_digital_async,you_digital_tmc}. These approaches introduce quantization, coding, or detection errors, but they improve compatibility with packet-based systems and can be more robust to phase asynchrony.}

\subsection{{Synchronization Levels used in this Article}}

{The AirFL mechanisms discussed in the following sections have different synchronization requirements. To avoid ambiguity, we use three synchronization levels throughout the rest of the article.}

{\textit{Fine synchronization} refers to frame and symbol-level timing alignment together with tight carrier-frequency and carrier-phase stability. This level is needed for phase-sensitive transmitter-side compensation, channel inversion, and joint transceiver designs in which the effective channel used for precoding, feedback, or coherent aggregation must remain valid during transmission.}

{\textit{Intermediate synchronization} refers to frame and symbol-level alignment together with partial phase stability. This level is sufficient for schemes that do not require exact phase cancellation but still need the residual phase error to remain within a controlled range, such as quadrant phase compensation.}

{\textit{Coarse synchronization} refers to round/frame alignment and sufficient waveform overlap, possibly supported by guard intervals, cyclic prefixes, receiver equalization, statistical averaging, or noncoherent/digital processing. Coarse synchronization does not require transmitter-side phase coherence, but it should not be interpreted as fully uncoordinated transmission: the received waveform must still preserve a common computation window and the correspondence among the model symbols being aggregated.}

\section{CSIT-Aware Over-the-Air Federated Learning}

CSIT-aware AirFL refers to schemes in which channel information is used for transmitter-side pre-compensation before wireless superposition. Following the CSI terminology introduced in the AirComp fundamentals, we discuss two representative variants: local-CSIT AirFL and global-CSIT AirFL. In both cases, CSIT is leveraged through transmission \textit{power control}, which serves as the key design element of this class.

\subsection{Local CSIT AirFL}

{The representative local-CSIT approach discussed in this subsection is developed for scalar effective channels, particularly a single-antenna server. Extending local-CSIT AirFL to multi-antenna servers is not a direct replacement of the scalar channel by a vector or matrix, because each device must design its transmit beamformer using only local channel knowledge while still contributing to a common aggregation function at the receiver.} Originally introduced in~\cite{khabin, localCSIT}, this approach has since been widely adopted in subsequent AirFL studies, including~\cite{gunduz_bpsk, khabin_sg, khabin_noise, azimi_hiersg}, where it has been extended to support digital modulations through one-bit aggregation~\cite{gunduz_bpsk}, large-scale cellular deployments~\cite{khabin_sg}, hierarchical multi-cluster networks~\cite{azimi_hiersg}, and principal-component-based learning~\cite{khabin_noise}. It comprises the following key components:

\textbf{Power Control:}
The following power control at each device $k$ employs direct channel compensation to exclusively align transmissions with the aggregation function~\eqref{simple_agg}.
\begin{align}
	p_k \propto \frac{1}{h_k} = \frac{1}{|h_k|}e^{-j\angle h_k},
\end{align}
where $h_k$ is the channel between device $k$ and the server.
However, this results in extremely high transmit power when the channel gain $|h_k|$ is small. To address this, \textit{truncated power control}, where only devices with sufficiently strong channels transmit, is used as
\begin{align}
	\label{powerall}
	p_k =
	\begin{cases}
		\frac{\sqrt{\rho}}{|h_k|}e^{-j\angle h_k}, & \text{if } |h_k|^2 \geq \theta, \\
		0, & \text{otherwise},
	\end{cases}
\end{align}
where $\rho$ is a denormalizing factor and $\theta$ denotes a threshold.

{This channel-inversion rule is phase-sensitive and therefore requires fine synchronization as defined in Section~\ref{sec:synchronization_airfl}. The effective scalar channel used in~\eqref{powerall} must include any residual timing, carrier-frequency, sampling-frequency, and phase offsets, and it must remain stable between channel estimation and model transmission. Otherwise, the inversion may no longer align the transmitted model symbols at the server.}

Accordingly, the active device set, which includes the transmitting devices, is defined as
\begin{align}
	\label{localselection}
	\mathcal{S} = \left\{k \mid |h_k|^2 \geq \theta \right\}.
\end{align}
Thus, \textit{device selection} is inherent in this approach.

\textbf{Aggregation:}
The single-antenna server estimates the aggregation function as
\begin{align}
	\mathbf{w}_\text{G} = \frac{\mathbf{y}}{\sqrt{\rho}|\mathcal{S}|}.
\end{align}
The result is unbiased and can be written as
\begin{align}
	\mathbf{w}_\text{G} 
	&= 
	\underbrace{\frac{1}{|\mathcal{S}|} \sum_{k \in \mathcal{S}} \mathbf{w}_k}_{\textcolor{darkgreen}{\text{Desired Aggregation}}}
	+ 
	\underbrace{\frac{\mathbf{z}}{\sqrt{\rho}|\mathcal{S}|}}_{\textcolor{red}{\text{Error from AWGN}}}.
\end{align}
The factor $\rho$ controls the error in recovering the aggregation $\frac{1}{|\mathcal{S}|} \sum_{k \in \mathcal{S}} \mathbf{w}_k$, and maximizing it leads to improved performance. 

\textbf{Denormalizing:} Assuming Rayleigh fading, i.e., $|h_k|^2 \sim \exp(1)$, the factor $\rho$ is chosen under the power constraint $P$ as
\begin{align}
	\mathbb{E}\left[|p_k|^2\right] = \rho  \mathbb{E}\left[\frac{1}{|h_k|^2} \,\middle|\, |h_k|^2 \geq \theta \right] = \rho  \text{Ei}(\theta) = P.
\end{align}
Thus,
\begin{align}
	\rho = \frac{P}{\text{Ei}(\theta)},
\end{align}
where $\text{Ei}(x) = \int_{x}^{\infty}\frac{e^{-t}}{t} \mathrm{d}t$ is the exponential integral function.

\textbf{Limitations:}
\begin{itemize}
	\item {It requires fine synchronization and fresh local effective CSI, because scalar channel inversion is valid only when the estimated effective channel remains stable over the aggregation block.}
	\item {It is tied to scalar channel inversion with single-antenna devices and a single-antenna server; hence it does not exploit transmit or receive spatial degrees of freedom.}
	\item It requires careful device selection and consequently limits participation, being highly dependent on channel conditions, which may introduce bias and adversely affect learning fairness; in extreme cases, no device may participate, leading to learning failure.
\item {It overlooks the learning aspect in its design, as the representative channel-inversion method discussed in this subsection is mainly communication-driven and does not explicitly use a learning-convergence criterion to guide the design. This limitation, however, should not be interpreted as applying to all local-CSIT AirFL approaches, since convergence-aware designs have also been studied~\cite{sery_hetero}.}
\end{itemize}

\subsection{Global CSIT AirFL}
This approach adopts a centralized design and leverages an arbitrary number of antennas at the server. While it was originally developed in~\cite{globalCSIT}, subsequent AirFL studies~\cite{bereyhi, latif, ng} have continued to build on its framework---either by reducing its complexity~\cite{bereyhi, ng} or by extending it to other wireless settings, such as systems incorporating intelligent reflecting surface (IRS)~\cite{latif}. The key components of this approach are as follows:

\textbf{Normalization:} Each device normalizes its model vector as
\begin{align}
	\label{normalize}
	\bar{\mathbf{w}}_k = \frac{\mathbf{w}_k - \eta_k \mathbf{1}}{\sigma_k},
\end{align}
where
\[
\eta_k = \frac{1}{s} \sum_{i=1}^{s} w_{k,i}, \quad \sigma_k^2 = \frac{1}{s} \sum_{i=1}^{s}(w_{k,i} - \eta_k)^2,
\]
and transmits $p_k\bar{\mathbf{w}}_k$. The pair $(\eta_k, \sigma_k)$ is also shared with the server.

Normalizing the model parameters provides two key benefits for the subsequent aggregation estimator. First, zero-mean entries ensure that the estimator is unbiased. Second, unit-variance entries make the power of the estimation error independent of the specific values of the model parameters. This normalization ultimately enables MSE-based design.

\textbf{Aggregation:}
The multiple-antenna server computes its estimate based on the received signal $\mathbf{y}$ as
\begin{align}
	\label{globalestimate}
	\mathbf{w}_\text{G} = \bar{\sigma} \frac{\mathbf{b}^H \mathbf{y}}{\sqrt{\rho}|\mathcal{S}|} + \bar{\eta}\mathbf{1},
\end{align}
where $|\cal S|$ is the number of active devices in this approach, $\rho$ is the denormalizing factor, and
\[
\bar{\sigma} = \frac{1}{|\mathcal{S}|} \sum_{k \in \mathcal{S}} \sigma_k, \quad \bar{\eta} = \frac{1}{|\mathcal{S}|} \sum_{k \in \mathcal{S}} \eta_k,
\]
represent the averages over the active devices, ensuring an unbiased estimation. Then,~\eqref{globalestimate} can be written as
\begin{align}
	\label{expandglobal}
	\mathbf{w}_\text{G} 
	&= 
	\underbrace{\frac{1}{|\mathcal{S}|} \sum_{k \in \mathcal{S}} \mathbf{w}_k}_{\textcolor{darkgreen}{\text{Desired Aggregation}}}
	+ 
	\underbrace{\frac{\bar{\sigma}}{|\mathcal{S}|}\sum_{k \in \mathcal{S}} 
		\left( \frac{\mathbf{b}^H \mathbf{h}_k p_k}{\sqrt{\rho}} - 1 \right)\bar{\mathbf{w}}_k}_{\textcolor{orange}{\text{Error from Channel Misalignment with Aggregation Weights}}}
	+ 
	\underbrace{\frac{\bar{\sigma}\mathbf{b}^H \mathbf{z}}{\sqrt{\rho}|{\cal S}|}}_{\textcolor{red}{\text{Error from AWGN}}}.
\end{align}

\textbf{Power Control and Denormalizing:} To minimize MSE caused by the error terms in~\eqref{expandglobal}, the optimal $p_k$ is~\cite{globalCSIT}
\begin{align}
	\label{powersyn}
	p_k = \sqrt{\rho} \frac{(\mathbf{b}^H \mathbf{h}_k)^H}{|\mathbf{b}^H \mathbf{h}_k|^2}.
\end{align}

{The precoder in~\eqref{powersyn} and the receive equalizer $\mathbf{b}$ are designed from the effective channels observed at the server. Hence, global-CSIT AirFL requires fine synchronization and is sensitive to channel aging across the pilot-estimation, centralized-optimization, feedback, and model-transmission stages. Residual timing, carrier-frequency offset, sampling-frequency offset, or phase drift changes the effective channels and can invalidate the coherent combining intended by~\eqref{powersyn}.}

With power constraint $|p_k|^2 \leq P$, the optimal $\rho$ is~\cite{globalCSIT}
\begin{align}
	\rho = \min_{k \in \mathcal{S}} P |\mathbf{b}^H \mathbf{h}_k|^2,
\end{align}

which results to
\begin{align}
	\text{MSE} = \frac{\sigma_z^2}{P}  \max_{k \in \mathcal{S}} \frac{\|\mathbf{b}\|^2}{|\mathbf{b}^H \mathbf{h}_k|^2}.
\end{align}
If the sole objective were to compute the most accurate aggregation, one could simply minimize the MSE with respect to the remaining unknowns---the set of active devices $\cal S$ and the equalization vector $\mathbf{b}$. However, the primary objective is to enable learning; thus, the focus shifts to evaluating the convergence performance of the learning process.

{We next present one representative global-CSIT framework in which the convergence bound is explicitly used to guide device selection and equalization.}

\textbf{Convergence Analysis:}
The following key widely-used assumptions are made to facilitate the analysis.

\textit{Assumption 1 (Lipschitz-Continuous Gradient):} The gradient of the loss function $F(\mathbf{w})$, as represented in~\eqref{lossfunction}, is characterized by Lipschitz continuity with a non-negative constant $L > 0$. This implies that for any pair of model vectors $\mathbf{w}_1$ and $\mathbf{w}_2$, we have
\begin{align}
	&F(\mathbf{w}_2) \leq F(\mathbf{w}_1) + \nabla F(\mathbf{w}_1)^T (\mathbf{w}_2-\mathbf{w}_1) + \frac{L}{2} \|\mathbf{w}_2 - \mathbf{w}_1\|^2.
\end{align}


\textit{Assumption 2 (Gradient Variance Bound):} The local stochastic gradient estimate for device $k$ at $\mathbf{w}_k$, using a mini-batch $\boldsymbol \xi_k$ with size $B$, is an unbiased estimate
of the ground-truth gradient $\nabla F(\mathbf{w}_k)$ with bounded variance
\begin{align}
	\mathbb{E}\left[\|\nabla F_k(\mathbf{w}_{k}, \boldsymbol\xi_k) - \nabla F(\mathbf{w}_k)\|^2\right] \leq \frac{\sigma_\text{g}^2}{B}.
\end{align}

\begin{tcolorbox}[colframe=orange!80!black, colback=orange!10, coltitle=white, title=\textbf{Convergence of Global CSIT AirFL}]
	Let $1 - \frac{L^2 \mu^2}{2} \tau (\tau - 1) - L \mu \tau \geq 0$. Under Assumptions 1 and 2, the convergence rate is bounded as~\cite{wafel}
	\begin{align}
		\label{conv_global}
		\frac{1}{T} \sum_{t=0}^{T-1} \mathbb{E}\left[ \|\nabla F(\mathbf{w}_{\text{G}}^t)\|^2 \right] 
		&\leq \frac{2(F(\mathbf{w}_{\text{G}}^0) - F(\mathbf{w}^*))}{\mu \tau T} 
		+ \frac{L^2 \mu^2 (\tau - 1)}{2} \frac{\sigma_\text{g}^2}{B} 
		+ \frac{L}{\mu \tau T} \sum_{t=0}^{T-1} \mathcal{I}_t,
	\end{align}
	where
	\begin{align}
		\mathcal{I}_t = \mu^2  \frac{\sigma_\text{g}^2}{B}  \frac{\tau}{|{\cal S}_t|} + \text{MSE}_t.
	\end{align}
\end{tcolorbox}

The bound in \eqref{conv_global} shows that the average squared gradient along the trajectory decomposes into three contributions. 
The first term, $\frac{2(F(\mathbf{w}_{\text{G}}^0) - F(\mathbf{w}^*))}{\mu \tau T}$, is a \emph{transient} term: it decays as $1/T$ and captures how far the algorithm starts from the optimum. 
The second term, $\frac{L^2 \mu^2 (\tau - 1)}{2} \frac{\sigma_\text{g}^2}{B}$, is an \emph{irreducible noise floor} due to stochastic gradients: larger mini-batches $B$ and smaller step sizes $\mu$ reduce this term, while too many local steps $\tau$ amplify it. 
The third term, $\frac{L}{\mu \tau T} \sum_{t=0}^{T-1} \mathcal{I}_t$, captures the \emph{impact of wireless imperfections}: $\mathcal{I}_t$ grows when fewer devices participate (small $|{\cal S}_t|$) or when the MSE is large. 
Thus, good convergence requires (i) enough rounds $T$, (ii) a carefully chosen learning rate and number of local steps, and (iii) system designs that keep the MSE small while involving as many devices as possible.

\textbf{Device Selection and Equalization:}
One approach to optimizing the convergence rate in~\eqref{conv_global} is to maximize device participation while keeping the MSE below a specified threshold $\theta$, as 
\begin{align}
	\label{csit_opt}
	\max_{\mathcal{S}, \ \mathbf{b}} |\mathcal{S}|,
\end{align}
subject to
\begin{align}
	\max_{k \in \mathcal{S}} \frac{\|\mathbf{b}\|^2}{|\mathbf{b}^H \mathbf{h}_k|^2} \leq \theta.\nonumber
\end{align}
This is a mixed combinatorial and non-convex optimization problem. Two main approaches have been proposed to solve~\eqref{csit_opt}.

\begin{itemize}
	
	\item \textit{Difference-of-Convex (DC) Programming Approach~\cite{globalCSIT}:} This approach tackles~\eqref{csit_opt} as a sparse and low-rank optimization problem. It employs DC programming to induce sparsity in device selection and enforce a rank-one constraint essential for effective equalization. The main advantage of this approach lies in its theoretical rigor and guaranteed global convergence. However, it comes with a high computational complexity of order $\mathcal{O}\Big(K\big((M^{2}+K)^{3} + M^{6}\big)\Big)$, which can be a limitation in large-scale or time-sensitive deployments.
	
	\item \textit{Matching Pursuit (MP) Based Greedy Scheduling Approach~\cite{bereyhi}:} This approach reformulates~\eqref{csit_opt} as a sparse support selection problem and solves it using a greedy, iterative matching pursuit algorithm inspired by compressive sensing. By iteratively selecting devices that contribute least to the MSE, it achieves a near-optimal solution with significantly lower computational burden. Its complexity scales polynomially with the number of devices and antennas as  $\mathcal{O}\left(K^{2}M^{2}\right)$, making it efficient and scalable. However, the approach is heuristic in nature and lacks the theoretical optimality guarantees of DC programming.
	
\end{itemize}
The above optimization flow of Global CSIT AirFL can be represented by three sequential stages:
\par\vspace{0.9em}
\begin{tikzpicture}[
	scale=1.1,
	transform shape,
	node distance=1.3cm,
	block/.style={
		rectangle,
		draw,
		rounded corners,
		thick,
		minimum width=3.4cm,
		minimum height=0.9cm,
		text centered,
		fill=gray!10,
		font=\sffamily\footnotesize
	},
	arrow/.style={-{Latex[length=2.2mm,width=1.6mm]}, thick}
	]
	\node[block, fill=green!10] (block1) {Joint Device Selection \& Equalization};
	\node[block, fill=yellow!10, right=of block1] (block2) {Denormalizing};
	\node[block, fill=orange!10, right=of block2] (block3) {Power Control};
	
	\draw[arrow] (block1) -- (block2);
	\draw[arrow] (block2) -- (block3);
	
\end{tikzpicture}

\textbf{Limitations:}
\begin{itemize}
\item {It requires global effective CSI at the edge server and reliable feedback of device-specific precoding coefficients, creating significant signaling overhead and channel-aging sensitivity.}

\item {Residual timing, carrier-frequency, sampling-frequency, or phase errors can mismatch the optimized precoders and equalizer to the actual aggregation channel, thereby invalidating the coherent combining implied by~\eqref{powersyn}.}

	\item It has high computational complexity due to the joint optimization~\eqref{csit_opt}.
	\item Power control may exceed device capabilities, particularly in low-power IoT applications.
	\item Owing to transmission power constraints, device selection becomes a critical factor, and fairness can be adversely affected. Nevertheless, as this scheme prioritizes maximizing the number of participating devices in~\eqref{csit_opt}, it seeks to maintain high learning performance despite these constraints.
	
\end{itemize}

\section{Blind Over-the-Air Federated Learning}

This approach avoids CSIT-based power control by allowing devices to transmit model updates at a constant power level, regardless of their channel magnitudes, a feature commonly referred to as \textit{blindness}. Notably, this strategy is in harmony with traditional wireless system designs, where constant-power, channel-agnostic transmission is widely used. In contrast, CSIT-aware AirFL introduces additional complexity due to the need for accurate transmitter-side channel adaptation and dynamic power control. Beyond simplifying system requirements, blind transmission offers several practical advantages: it maintains average energy consumption regardless of channel variations, avoids expanding the dynamic range of transmitted signals---thereby simplifying hardware and reducing costs---and reduces the risk of performance degradation from inaccurate pre-compensation. An additional advantage is that all devices can contribute to the aggregation process, thereby maximizing the inclusion of locally trained models in the global learning update. 

Two main variants exist within this approach: fully blind AirFL, which operates without CSIT, and partial-phase-aware blind AirFL, where each device has access only to limited channel phase information. The fully blind approach emphasizes receiver-side post-processing and high-dimensional \textit{equalization} as its key design element, whereas the partial-phase-aware approach uses limited phase compensation to prevent destructive sign inversions.

\subsection{Fully Blind AirFL}
The server, equipped with a {sufficiently} large number of antennas, leverages high dimensional statistical properties to aggregate effectively. This approach was initially proposed in~\cite{blind} and subsequently examined in greater depth in~\cite{saeed, turky_adc, turky_time, hier_deniz}, providing deeper analyses~\cite{saeed} and several extensions to more practical and complex scenarios, including hardware-impaired transceivers~\cite{turky_adc}, hierarchical multi-cluster architectures~\cite{hier_deniz}, and time-varying fading environments~\cite{turky_time}. The main process consists of the following key components:

\textbf{Signal Assumption:} No pre-processing is performed at the devices. Assuming a constant transmit power $P$ and coarse synchronization as defined in Section~\ref{sec:synchronization_airfl}, the signal received at the $m$-th antenna of the server can be expressed, based on~\eqref{MACsig}, as
\begin{align}
	\mathbf{y}_m = \sum_{k=1}^{K} \sqrt{P}h_{k,m} \mathbf{w}_k + \mathbf{z}_m.
\end{align}
Here, the channel coefficients are further assumed to be i.i.d. and Gaussian distributed, with  
\begin{align}
	\mathbb{E}\left[h_{k,m}\right] = 0, \quad \mathbb{E}\left[|h_{k,m}|^2\right] = \sigma_h^2, \forall k, m.
\end{align}
{This scheme requires waveform overlap of corresponding model coordinates. Residual timing offsets can be tolerated when they remain within the waveform tolerance or are mitigated by receiver-side processing. Residual carrier-frequency and phase offsets can be tolerated only when their induced rotations are sufficiently stable over the aggregation block or can be treated as part of the effective random channels. If fast phase drift or CFO makes the effective channels vary significantly across model coordinates, the antenna-domain averaging effect in~\eqref{blind_agg} is weakened.}

\textbf{Equalization and Aggregation:} The server applies a special high-dimensional equalization over all antennas as
\begin{align}
	\mathbf{b} = \left[\left(\sum_{k=1}^{K} h_{k,1}\right)^H, \ldots, \left(\sum_{k=1}^{K} h_{k,M}\right)^H \right],
\end{align}
and estimates as
\begin{align}
	\label{blind_agg}
	\mathbf{w}_\text{G} = \frac{\sum_{m=1}^{M}\left(\sum_{k=1}^{K}h_{k,m}\right)^H \mathbf{y}_m}{K M \sigma_h^2}.
\end{align}

\textbf{Aggregation Behavior as $M \to \infty$:} We now analyze how the aggregation converges to the true average as the number of antennas grows.
\begin{align}
	&\lim_{M \to \infty} \frac{\sum_{m=1}^{M}\left(\sum_{k=1}^{K}h_{k,m}\right)^H \left(\sum_{k=1}^{K}h_{k,m}\mathbf{w}_k + \mathbf{z}_m\right)}{K M\sigma_h^2} = \lim_{M \to \infty} \underbrace{\frac{1}{K} \sum_{k=1}^{K} 
		\left( \frac{\sum_{m=1}^{M} |h_{k,m}|^2 }{M\sigma_h^2} \right) 
		\mathbf{w}_k}_{\textcolor{darkgreen}{\text{Desired Aggregation}}} \nonumber \\
	&\quad + 
	\underbrace{\frac{\sum_{}^{} \sum_{k' \neq k}^{} \sum_{m=1}^{M} 
			(h_{k,m})^H h_{k',m} \mathbf{w}_{k'}}{K M\sigma_h^2}}_{\textcolor{orange}{\text{Error from Interference}}} \quad + 
	\underbrace{\frac{\sum_{m=1}^{M} 
			\left(\sum_{k=1}^{K} h_{k,m} \right)^H \mathbf{z}_m}{K M\sigma_h^2}}_{\textcolor{red}{\text{Error from AWGN}}} =   \frac{1}{K} \sum_{k=1}^{K} \mathbf{w}_k.
\end{align}

Here, the result relies on the large-number theory as
\begin{align}
	\frac{\sum_{m=1}^{M} |h_{k,m}|^2 }{M} \to \sigma_h^2,
\end{align}
which is a result of the i.i.d. channel assumption. The cross-terms due to interference and noise contributions vanish as $M \to \infty$.

\textbf{Number of Antennas for Convergence:} A probabilistic lower bound on the number of antennas $M$ required to ensure a bounded estimation error in~\eqref{blind_agg} is established. Specifically, the following theorem shows that, for sufficiently large $M$, the estimation error remains bounded with high probability, thereby guaranteeing convergence.
\begin{tcolorbox}[colframe=red, colback=red!10, coltitle=white, title=\textbf{Minimum Number of Antennas for Convergence}]
	The absolute and expected estimation error in~\eqref{blind_agg} are bounded as~\cite{saeed}
	\begin{align}
		\left\|\mathbf{w}_\text{G} -\frac{1}{K} \sum_{k=1}^{K}\mathbf{w}_k\right\| &\leq \frac{\epsilon}{K}, \\
		\mathbb{E}\left[\left\|\mathbf{w}_\text{G} -\frac{1}{K} \sum_{k=1}^{K}\mathbf{w}_k\right\|\right] &\leq \frac{4\gamma_n}{\sqrt{M} c_n} \left( \sqrt{\pi} + \ln(6K) \right),
	\end{align}
	where $c_n = {1}/{\gamma_n} + {\sigma_h}/{\sigma_z}$ and $\gamma_n$ is a positive constant. Moreover, to satisfy the above with probability at least $1 - \delta$, the number of antennas $M$ must fulfill
	\begin{align}
		M \geq \frac{8\gamma_n^2 K^2}{\epsilon^2 c_n^2} \ln \left( \frac{6K}{\delta} \right).
	\end{align}
\end{tcolorbox}

The error $\|\mathbf{w}_\text{G} - \frac{1}{K}\sum_{k}\mathbf{w}_k\|$ scales roughly as $1/\sqrt{M}$: more antennas average out interference and noise more effectively. 
The parameter $\epsilon$ controls how tight the approximation must be (smaller $\epsilon$ means stricter accuracy and thus larger $M$), while $\delta$ specifies the confidence level (smaller $\delta$ means the guarantee must hold with higher probability, again requiring more antennas). 
The dependence $M \propto K^2 \ln(6K/\delta)$ shows that, as the number of devices grows, the receiver needs more antennas to maintain the same aggregation quality. 
In short, this bound formalizes the \textit{massive-MIMO averaging} effect: with sufficiently many antennas, blind AirFL can approximate exact averaging arbitrarily well, but the required array size grows with both network size and desired reliability.

\textbf{Limitations:}
\begin{itemize}
	\item {The antenna array required to reach a target aggregation accuracy can be large, especially as the number of devices or reliability requirements increase, which may be costly or infeasible in some deployments.}
	\item The primary assumption underlying this approach is the presence of i.i.d. channel conditions; however, this assumption does not always hold in practical wireless environments.
	\item {It relaxes transmitter-side phase synchronization, but still requires waveform overlap; fast CFO or phase drift can weaken the aggregation effect.}
	\item No optimization or adaptation is incorporated in this design, rendering it suboptimal; however, this simplification has been made to preserve the tractability and analytical clarity of the design.
\item {With finite antenna arrays, residual interference and noise remain after spatial averaging; if these errors are large relative to the learning tolerance, they can degrade convergence, final accuracy, and fairness.}
\end{itemize}

\subsection{Partial-Phase-Aware Blind AirFL} 	

This blind approach is primarily applicable in the single-antenna setting and assumes that each device~$k$ has access to a partial estimate of its channel phase~$\angle h_k$, denoted by~$\angle_\text{p} h_k$, such that the residual phase error satisfies $\left|\angle h_k-\angle_\text{p} h_k\right|<\frac{\pi}{2}$ after phase wrapping. Using this partial information, each device performs \textit{quadrant phase compensation}, adjusting its transmission phase so that the real part of the effective channel remains positive at the server. This technique accommodates a wide inaccuracy range---covering one-quarter of the entire phase space---and prevents channel-induced sign inversions in the aggregated signal. {It therefore requires intermediate synchronization as defined in Section~\ref{sec:synchronization_airfl}: exact phase cancellation is not needed, but the residual phase error must remain within the compensated quadrant during one aggregation block; otherwise, some effective channel signs may be flipped, leading to biased aggregation.}

No additional pre-processing is required at the device side, and due to the inherent limitations of the single-antenna scenario, post-processing at the receiver is not applied either. Initially proposed in~\cite{blind_simple1}, later enhanced with momentum-based acceleration in~\cite{blind_simple2}, and further extended in~\cite{blind_interference}, this scheme demonstrates that AirFL can achieve convergence even without sophisticated signal processing. Although performance is degraded, this method serves as a useful lower-bound benchmark for AirFL systems under minimal system complexity.

{The generalized interference model is discussed in this subsection to better understand the intrinsic behavior of AirFL under general wireless disturbance conditions, beyond the usual additive Gaussian noise assumption. The representative analysis in~\cite{blind_interference} models electromagnetic interference through a heavy-tailed distribution and provides an explicit convergence guarantee under this generalized disturbance. This is particularly informative in the partial-phase-aware blind setting because the system does not force the received signal toward an ideal average through sophisticated transmitter-side pre-processing or receiver-side equalization. Instead, the wireless superposition itself determines the imposed aggregation structure, allowing one to see how learning behaves under more natural and imperfect aggregation conditions. This should not be interpreted as a unique ability of the partial-phase-aware blind setting to cope with interference. Similar interference models could also be incorporated into CSIT-aware, fully blind, or weighted AirFL frameworks, but the resulting convergence behavior would depend on the specific aggregation structure and optimization design.}

{\textbf{Signal Model:} Following this generalized disturbance model, the interference vector $\boldsymbol{\xi}$ is used in place of the AWGN term $\mathbf{z}$. Accordingly, the real part of the received signal at the single-antenna server is expressed as}
\begin{align}
	\mathbf{y}_\text{r} = \sum_{k=1}^{K} \sqrt{P} h_{\text{r},k} \mathbf{w}_{k} + \boldsymbol{\xi},
\end{align}
where $h_{\text{r},k} = h_k \cos(\angle h_k - \angle_\text{p} h_k)$ is the real part of channel $h_k$ after phase compensation.

Two critical assumptions are made:

- All channels $\{h_{\text{r}, k}\}$ are i.i.d. random variables with mean $\omega$ and variance $\sigma^2$.

- The interference vector $\boldsymbol{\xi}$ has i.i.d. entries following a symmetric $\alpha$-stable distribution.

It has been well established, both theoretically and empirically, that electromagnetic interference in wireless systems often follows a heavy-tailed distribution~\cite{tailtheory, tailexperiment}. To capture this behavior, the interference can be modeled using the symmetric $\alpha$-stable distribution, defined below.

\textit{Definition:}
A random variable $\xi$ is said to follow a symmetric $\alpha$-stable distribution if its characteristic function is given by
\begin{align}
	\mathbb{E}\left[e^{j\omega \xi}\right] = \exp(-\delta^\alpha |\omega|^\alpha),
\end{align}
{where $\delta > 0$ is the scale parameter and $\alpha \in (0,2]$ is the tail index. The tail index controls the impulsiveness of the interference: $\alpha=2$ corresponds to the Gaussian case, $\alpha=1$ corresponds to the Cauchy case, and smaller values of $\alpha$ represent heavier-tailed interference.}

\textbf{Aggregation:} The aggregated signal without any post-processing is estimated as
\begin{align}
	\mathbf{w}_\text{G} = \frac{\mathbf{y}}{K\sqrt{P}},
\end{align}
which is equal to
\begin{align}
	\mathbf{w}_\text{G} 
	&= 
	\underbrace{\frac{1}{K} \sum_{k=1}^{K} h_{\text{r}, k} \mathbf{w}_k + \frac{\boldsymbol{\xi}}{K\sqrt{P}}}_{\textcolor{violetblue}{\text{Imposed Aggregation}}}.
\end{align}
Here, the desired aggregation function is not explicitly defined; instead, the actual outcome, including all its imperfections, is regarded as the imposed aggregation function.

\textbf{Convergence Analysis:} Surprisingly, even without any form of signal processing (except phase compensation), the partial-phase-aware blind scheme can still converge despite the presence of heavy-tailed interference. To support the analysis, the following assumptions are introduced in addition to Assumptions 1 and 2:

\textit{Assumption 3 (Strong Convexity):} 
The loss function $F(\mathbf{w})$ is $\gamma$-strongly convex, i.e., for any $\mathbf{w}_1$ and $\mathbf{w}_2$, it satisfies
\begin{align}
	F(\mathbf{w}_2) 
	\geq F(\mathbf{w}_1) 
	+ \nabla F(\mathbf{w}_1)^{\top} (\mathbf{w}_2 - \mathbf{w}_1) 
	+ \frac{\gamma}{2} \|\mathbf{w}_2 - \mathbf{w}_1\|^2.
\end{align}

\textit{Assumption 4 (Positive Definiteness of the Hessian):} 
For any given vector $\mathbf{w}$, the Hessian matrix of 
$F(\mathbf{w})$, i.e., $\nabla^2 F(\mathbf{w})$, is $\alpha$-positive definite. 
A symmetric matrix $\mathbf{Q}$ is said to be $\alpha$-positive definite if 
\[
\mathbf{v}^{\top} \mathbf{Q} \mathbf{v}^{\langle \alpha - 1 \rangle} > 0, 
\quad \forall\, \mathbf{v} \text{ with } \|\mathbf{v}\|_{\alpha}^{\alpha} > 1,
\]
where the operator $\mathbf{v}^{\langle \alpha - 1 \rangle}$ denotes the 
\emph{signed-power vector} defined element-wise as
\[
\mathbf{v}^{\langle \alpha - 1 \rangle}
= \big[\operatorname{sgn}(v_1)|v_1|^{\alpha - 1},
\operatorname{sgn}(v_2)|v_2|^{\alpha - 1},
\ldots,
\operatorname{sgn}(v_d)|v_d|^{\alpha - 1}\big]^{\top},
\]
and the $\alpha$-norm is defined as
\[
\|\mathbf{v}\|_{\alpha}^{\alpha}
= \left( \sum_{i=1}^{d} |v_i|^{\alpha} \right)^{\!1/\alpha}.
\]

\textit{Assumption 5 (Bounded Gradient):} 
The gradient at each device $k$ is bounded as
\begin{align}
	\|\nabla F_k(\mathbf{w})\| \leq G,
\end{align}
for all $k$ and all model parameters $\mathbf{w}$.

\textit{Assumption 6 (Bounded Interference):}
The $\alpha$-moment of the interference is bounded as
\begin{align}
	\mathbb{E}\left[\|\boldsymbol{\xi}\|_\alpha^\alpha\right] \leq C,
\end{align}
where $C > 0$ is a constant.

\begin{tcolorbox}[colframe=red, colback=red!10, coltitle=white, title=\textbf{Convergence of Partial-Phase-Aware Blind AirFL}]
	If the learning rate in~\eqref{update_step} is set to diminish as $\mu_t = \theta/t$ where $\theta > \frac{\alpha - 1}{\omega L}$, then the scheme under Assumptions 1--6 converges as~\cite{blind_interference}
	\begin{align}
		\label{con_tail}
		\mathbb{E} \left[\| \mathbf{w}_\text{G}^t - \mathbf{w}^* \|_\alpha^\alpha \right]
		&\leq \frac{4\theta^\alpha \left( C + \sigma^\alpha G^\alpha s^{1 - \frac{1}{\alpha}} / K^{\alpha/2} \right)}{\omega \theta L - \alpha + 1} \cdot \frac{1}{t^{\alpha - 1}}.
	\end{align}
\end{tcolorbox}
{The bound in~\eqref{con_tail} provides the AirFL-specific significance of the $\alpha$-stable interference model. The tail index $\alpha$ does not only describe the shape of the interference distribution; it directly determines the learning-error decay rate. When $\alpha=2$, corresponding to the Gaussian case, the rate becomes order $1/t$. As $\alpha$ decreases, the interference becomes more impulsive and the rate slows to order $1/t^{\alpha-1}$. Therefore, heavier-tailed wireless interference directly slows the convergence of partial-phase-aware blind AirFL. The constant in front also captures the effect of the interference strength $C$, the channel randomness $\sigma$, the gradient bound $G$, the model dimension $s$, and the number of devices $K$. Thus, the interference environment affects AirFL not only as an aggregation disturbance, but also through the final learning convergence behavior.}

\textbf{Limitations:}
\begin{itemize}
	\item The design is inherently rigid, offering no flexibility for optimization or adaptation.
	\item {Quadrant phase compensation requires intermediate synchronization: frame/symbol alignment and carrier-frequency/phase stability sufficient to keep the residual phase error within the compensated quadrant.}
	\item It heavily relies on restrictive channel, interference, and learning assumptions.
	\item Since the aggregation weights are simply the channel coefficients, this leads to biased aggregation and fairness issues, resulting in degraded performance.
\end{itemize}

\section{Weighted Over-the-Air Federated Learning} 

Weighted AirFL, also known as WAFeL, offers a flexible approach that avoids CSIT-based power control and large antenna arrays by introducing and emphasizing a new design dimension: \textit{aggregation weights}. This enables operation even with a single-antenna server. This approach was originally proposed in~\cite{wafel, wafel_conf}, encompassing both homogeneous and heterogeneous computational settings, and was subsequently extended in~\cite{w_japan} to regression problems, with the following key components:

\textbf{Weighted Aggregation:} Other AirFL approaches, like traditional FL, use fixed equal-weight aggregation as in~\eqref{simple_agg}. This simple aggregation is grounded on ideal communication conditions, yet remains adopted despite interference, noise, and other wireless impairments. In such cases, AirComp aims to approximate the fixed aggregation function as closely as possible, with the accuracy of this approximation depending on the effectiveness of the system design. To overcome this, WAFeL proposes a general weighted aggregation as
\begin{align}
	\mathbf{w}_{\text{G}}^t  = {\sum_{k=1}^{K} \alpha_{k}^t \mathbf{w}_{k}^t},
\end{align}
where $\alpha_{k}^t \geq 0$ is the weight corresponding to device $k$, and the weight vector $\boldsymbol{\alpha}_t = [\alpha_{1}^t,\ldots,\alpha_{K}^t]^\top$, such that $\mathbf{1}^\top \boldsymbol \alpha_t = 1$. Here, rather than forcing one to approximate the other, both AirComp and the aggregation function are allowed to adapt toward each other, enabling more accurate and robust computation with reduced system requirements.

\textbf{Quadrant Phase Compensation and Normalization:} The WAFeL approach can be applied regardless of the number of antennas at the server. However, in the single-antenna case, it requires quadrant phase compensation. After normalization, as described in~\eqref{normalize}, the transmitted signal with constant power~$P$ is given by
\begin{align}
	\mathbf{x}_{k} = \sqrt{P} e^{-j \angle_\text{p} h_k} \bar{\mathbf{w}}_{k},
\end{align}
enabling blind transmission.

{In the single-antenna setting, WAFeL inherits the intermediate synchronization requirement of quadrant phase compensation. This requirement is weaker than fine synchronization because exact phase cancellation and CSIT-based power control are not needed. However, the effective channel matrix $\mathbf{H}$ used in the weighted aggregation model below must remain stable enough for the optimized weights and equalizer in~\eqref{weightderror} to represent the intended weighted aggregate.}

\textbf{Signal Model:} Arranging the real and imaginary components of the received signal into a vector yields the following real-valued representation at the single-antenna server:
\begin{align}
	\mathbf{Y} = \sqrt{P}\mathbf{H}\bar{\mathbf{W}} + \mathbf{Z},
\end{align}
where $\bar{\mathbf{W}} = [\bar{\mathbf{w}}_1, \ldots, \bar{\mathbf{w}}_K]^\top$, and
\begin{align}
	\mathbf{Y} = \begin{bmatrix}
		\mathfrak{Re}\left\{\mathbf{y^\top}\right\} \\
		\mathfrak{Im}\left\{\mathbf{y^\top}\right\}
	\end{bmatrix},\ 
	\mathbf{H} = \begin{bmatrix}
		\mathfrak{Re}\left\{\mathbf{h^\top}\right\} \\
		\mathfrak{Im}\left\{\mathbf{h^\top}\right\}
	\end{bmatrix},\ 
	\mathbf{Z} = \begin{bmatrix}
		\mathfrak{Re}\left\{\mathbf{z^\top}\right\} \\
		\mathfrak{Im}\left\{\mathbf{z^\top}\right\}
	\end{bmatrix},\nonumber
\end{align}
with $\mathbf{h} = [h_1 e^{j(\angle h_1 - \angle_\text{p} h_1)},\ldots,h_Ke^{j(\angle h_K - \angle_\text{p} h_K)}]^\top$.

\textbf{Aggregation:} The server employs an equalization vector $\mathbf{b} \in \mathbb{R}^{2 \times 1}$ to estimate the weighted aggregation as
\begin{align}
	\mathbf{w}_\text{G}^\top = \frac{1}{\sqrt{P}}\mathbf{b}^\top \mathbf{Y} + \sum_{k=1}^{K}\alpha_k\eta_k\mathbf{1}^\top,
\end{align}
which can be written as
\begin{align}
	\label{weightderror}
	\mathbf{w}_\text{G}^\top 
	&= 
	\underbrace{\sum_{k=1}^{K}\alpha_k\mathbf{w}_k^\top}_{\textcolor{darkgreen}{\text{Desired Aggregation}}}
	+ 
	\underbrace{\left(\mathbf{b}^\top \mathbf{H} - (\boldsymbol\alpha\odot\boldsymbol\sigma)^\top\right) \bar{\mathbf{W}}}_{\textcolor{orange}{\text{Error from Channel Misalignment with Aggregation Weights}}}
	+ 
	\underbrace{\frac{1}{\sqrt{P}}\mathbf{b}^\top \mathbf{Z}}_{\textcolor{red}{\text{Error from AWGN}}}.
\end{align}
where $\boldsymbol \sigma = [\sigma_1,\ldots,\sigma_K]^\top$ and $(\boldsymbol\alpha\odot\boldsymbol\sigma)^\top = [\alpha_1\sigma_1, \ldots,\alpha_K\sigma_K]$. Due to the normalized transmission, this estimation is unbiased.

\textbf{Equalization:} The optimal equalizer that minimizes MSE arising from the error terms in~\eqref{weightderror} is obtained as~\cite{wafel}
\begin{align}
	\mathbf{b}_\text{opt}^\top = (\boldsymbol\alpha\odot\boldsymbol\sigma)^\top \mathbf{H}^\top \left(\frac{\sigma_z^2}{P}\mathbf{I}_2+\mathbf{H}\mathbf{H}^\top\right)^{-1}.
\end{align}

This equalization leads to 
\begin{align}
	\text{MSE}(\boldsymbol\alpha) = s\boldsymbol\alpha^\top \text{diag}(\boldsymbol\sigma)\left( \mathbf{I}_K + \frac{P}{\sigma_z^2}\mathbf{H}^\top\mathbf{H} \right)^{-1}\text{diag}(\boldsymbol\sigma)\boldsymbol\alpha.
\end{align}
The final step involves selecting the aggregation weight vector $\boldsymbol \alpha$, which can be optimized to enhance convergence performance.

\textbf{Convergence Analysis:} The convergence rate of WAFeL is given in the next.
\begin{tcolorbox}[colframe=blue, colback=blue!10, coltitle=white,  title=\textbf{Convergence of WAFeL}]
	Let $1-\frac{L^2\mu^2}{2}\tau(\tau-1)-L\mu \tau \geq 0$ and $\boldsymbol \alpha_t$ be the weight vector at round $t$. Then the convergence rate under Assumptions 1 and 2 is bounded as~\cite{wafel_conf}
	\begin{align}
		\label{wafel_con}
		\frac{1}{T}\sum_{t=0}^{T-1}\mathbb{E}\left[\|\nabla F( \mathbf{w}_{\text{G}}^t)\|^2\right] &\leq \frac{2\left(F(\mathbf{w}_{\text{G}}^0)-F(\mathbf{w}^*)\right)}{\mu\tau T} + L^2\mu^2\frac{ \tau-1}{2}\frac{\sigma_\text{g}^2}{B} + \frac{L}{\mu \tau T}\sum_{t=0}^{T-1}{\cal I}_t(\boldsymbol \alpha_t),
	\end{align}
	where 
	\begin{align}
		{\cal I}_t(\boldsymbol \alpha_t) = \mu^2\frac{\sigma_\text{g}^2}{B}\tau {\|\boldsymbol \alpha_t\|^2} + \text{MSE}_t(\boldsymbol \alpha_t).
	\end{align}
\end{tcolorbox}
The convergence bound in \eqref{wafel_con} has the same qualitative structure as in the global CSIT case, but with an explicit dependence on the aggregation weights $\boldsymbol{\alpha}_t$. 
The first term, $\frac{2(F(\mathbf{w}_{\text{G}}^0)-F(\mathbf{w}^*))}{\mu \tau T}$, again captures the transient phase and decays as $1/T$. 
The second term, $\frac{L^2\mu^2(\tau-1)}{2}\frac{\sigma_\text{g}^2}{B}$, is the stochastic gradient noise floor. 
The third term, $\frac{L}{\mu \tau T}\sum_t {\cal I}_t(\boldsymbol \alpha_t)$, links convergence directly to the choice of weights and the wireless channel. 
Here, $ {\cal I}_t(\boldsymbol \alpha_t)$ has two parts: $\mu^2 \frac{\sigma_\text{g}^2}{B}\tau \|\boldsymbol\alpha_t\|^2$ reflects how unbalanced weights amplify gradient noise (large $\|\boldsymbol\alpha_t\|$ means a few devices dominate the update), while $\text{MSE}_t(\boldsymbol\alpha_t)$ quantifies the aggregation distortion due to AirComp. 
By optimizing $\boldsymbol\alpha_t$ to keep both $\|\boldsymbol\alpha_t\|^2$ and $\text{MSE}_t(\boldsymbol\alpha_t)$ small, WAFeL jointly balances \emph{learning fairness} and \emph{wireless reliability}, potentially leading to faster and more stable convergence than fixed, non-optimized weights.

\textbf{Aggregation Weight Selection:} The convergence rate in~\eqref{wafel_con} can be optimized by minimizing the norm of the weight vector while ensuring that the MSE remains below a threshold $\theta$, as
\begin{align}
	\label{weight_selection}
	\boldsymbol\alpha^t = \arg\min_{\boldsymbol\alpha \geq \mathbf{0}} \|\boldsymbol \alpha\|^2,
\end{align}
subject to
\begin{align}
	&\boldsymbol\alpha^\top \text{diag}(\boldsymbol\sigma_t)\left( \mathbf{I}_K + \frac{P}{\sigma_z^2}\mathbf{H}_t^\top\mathbf{H}_t \right)^{-1}\text{diag}(\boldsymbol\sigma_t)\boldsymbol\alpha \leq \theta,\nonumber \\
	&\mathbf{1}^\top \boldsymbol \alpha = 1.\nonumber
\end{align}

This is a convex optimization problem and can be efficiently solved using standard solvers. A low-complexity iterative algorithm with computational order $\mathcal{O}\left(K^{3}\right)$ for solving~\eqref{weight_selection} is presented in~\cite{wafel}.
This approach enables participation from all devices while adapting the aggregation weights to automatically account for both communication and learning aspects---the norm of the weight vector captures the learning contribution, whereas the MSE term reflects the communication quality. 

Therefore, the above optimization flow of WAFeL can be represented by two sequential stages: 

\par\vspace{0.8em}  
\hspace{2.5cm}
\begin{tikzpicture}[
	scale=1.1,
	transform shape,
	node distance=1.8cm,
	block/.style={
		rectangle,
		draw,
		rounded corners,
		thick,
		minimum width=3.6cm,
		minimum height=1.0cm,
		text centered,
		fill=gray!10,
		font=\sffamily\footnotesize
	},
	arrow/.style={-{Latex[length=2.2mm,width=1.6mm]}, thick}
	]
	\node[block, fill=cyan!15] (block1) {Weight Selection};
	\node[block, fill=violet!20, right=of block1] (block2) {Equalization};
	
	\draw[arrow] (block1) -- (block2);
	
\end{tikzpicture}

\textbf{Soft vs Hard Selection:}
The continuous flexibility in selecting aggregation weights in~\eqref{weight_selection}---referred to as the \textit{soft} device selection approach---offers lower complexity and greater potential for optimization. In contrast, the \textit{hard} device selection approaches used in local CSIT AirFL~\eqref{localselection} and global CSIT AirFL~\eqref{csit_opt} are limited to a small set of discrete choices. In particular, the global CSIT formulation involves a high-complexity integer optimization problem, further increasing the computational burden. Unlike hard selection, which forces a binary decision on whether a device is selected or not, the soft approach allows each device to contribute proportionally to its potential---even if that contribution is small---ensuring that all devices participate in the learning process.

\textbf{Extension to Computationally Heterogeneous Devices:} In heterogeneous environments, each device~$k$ possesses a unique hardware capability, typically characterized by its computing speed~$f_k$ (e.g., CPU or GPU). To meet a common computation deadline across all devices, the local batch size~$B_k$ is adjusted proportionally to~${1}/{f_k}$, effectively capturing computational heterogeneity. Under this setting, the heterogeneity-aware FedAvg aggregation is given by
\begin{align}
	\label{agg}
	\mathbf{w}_{\text{G}} = \frac{1}{B_\text{tot}} \sum_{k=1}^{K} B_k \mathbf{w}_{k},
\end{align}
where $B_\text{tot} = \sum_{k=1}^{K} B_k$, and the aggregation weights are compactly expressed as the vector $\mathbf{b}_\text{w} = \left[\frac{B_1}{B_\text{tot}}, \ldots, \frac{B_K}{B_\text{tot}} \right]^\top$.

The inequality introduced below is fundamental for quantifying computational heterogeneity through the batch sizes $B_k$, and forms the basis for the following convergence result.

\textit{Assumption 7 (Heterogeneity-Aware Gradient Variance Bound):} The local stochastic gradient estimate for device $k$ at $\mathbf{w}_k$, using a mini-batch $\boldsymbol \xi_k$ with size $B_k$, is an unbiased estimate
of the ground-truth gradient $\nabla F(\mathbf{w}_k)$ with bounded variance
\begin{align}
	\mathbb{E}\left[\|\nabla F_k(\mathbf{w}_{k}, \boldsymbol\xi_k) - \nabla F(\mathbf{w}_k)\|^2\right] \leq \frac{\sigma_\text{g}^2}{B_k}.
\end{align}
As the batch size increases, the computed gradient becomes more accurate.

\begin{tcolorbox}[colframe=blue, colback=blue!10, coltitle=white, title=\textbf{Convergence of Heterogeneity-Aware WAFeL}]
	Let $1-\frac{L^2\eta^2}{2}\tau(\tau-1)-L\eta \tau \geq 0$ and $\boldsymbol \alpha_t$ as the weight vector for each round $t$, then the convergence rate under Assumptions 1 and 7 is bounded as~\cite{wafel}
	\begin{align}
		\label{ratebound}
		&\frac{1}{T}\sum_{t=0}^{T-1}\mathbb{E}\left[\|\nabla F( {\mathbf{w}}_{\text{G}}^t)\|^2\right] \leq  \frac{2\left(F({\mathbf{w}}_{\text{G}}^0)-F^*\right)}{\eta \tau T}+\frac{L}{\eta \tau T}\sum_{t=0}^{T-1}{\cal I}_t(\boldsymbol \alpha_t),
	\end{align}
	where
	\begin{align}
		\label{Iterm}
		{\cal I}_t(\boldsymbol \alpha_t) &= {L\eta^3}\frac{\tau(\tau-1)}{2}{\sigma_\text{g}^2} {\boldsymbol \alpha_t}^\top \mathbf{b}_\text{s}+{ \eta^2}{\sigma_\text{g}^2}\tau {\boldsymbol \alpha_t}^\top \text{diag}\left\{\mathbf{b}_\text{s}\right\} \boldsymbol \alpha_t+ {\text{MSE}_t(\boldsymbol \alpha_t)},
	\end{align}
	where the vector $\mathbf{b}_\text{s} = [\frac{1}{B_1},\ldots,\frac{1}{B_K}]^\top$ is formed from the inverse of batch sizes of all the devices.
\end{tcolorbox}

In the heterogeneous setting, the asymmetry shows up explicitly in ${\cal I}_t(\boldsymbol\alpha_t)$ through the vector $\mathbf{b}_\text{s}$. 
Devices with smaller $B_k$ produce noisier gradients (larger $1/B_k$), which makes them more dangerous to overweight. 
The first two terms in \eqref{Iterm} describe how the combination of local steps, gradient noise, and the chosen weights interacts with this heterogeneity: large weights on low-batch devices can substantially increase ${\cal I}_t(\boldsymbol\alpha_t)$ and thus slow convergence. 
The third term, $\text{MSE}_t(\boldsymbol\alpha_t)$, again captures the AirComp distortion. 
The optimization problem in \eqref{practical_obj2} therefore seeks weights that down-weight unreliable (small-batch) devices just enough to control the noise, while also respecting an MSE constraint that ensures acceptable wireless aggregation quality. 
Intuitively, the bound shows that WAFeL can \textit{re-balance} the influence of fast and slow devices: it lets more reliable, high-batch devices have a stronger say in the global update, without completely ignoring weaker devices, thereby achieving convergence even under strong computational heterogeneity.

Therefore, the weight selection problem to optimize the convergence rate can be characterized as
\begin{align}
	\label{practical_obj2}
	\boldsymbol\alpha_t = \arg\min_{\boldsymbol\alpha \geq \mathbf{0}} \boldsymbol \alpha^\top \text{diag}\left\{\mathbf{b}_\text{s}\right\} \boldsymbol \alpha, 
\end{align}
subject to
\begin{align}
	&\boldsymbol\alpha^\top \text{diag}(\boldsymbol\sigma_t)\left( \mathbf{I}_K + \frac{P}{\sigma_z^2}\mathbf{H}_t^\top\mathbf{H}_t \right)^{-1}\text{diag}(\boldsymbol\sigma_t)\boldsymbol\alpha \leq \theta,\nonumber \\&\mathbf{1}^\top \boldsymbol{\alpha} = 1,\nonumber
\end{align}
which is a convex problem. Here, the term \( \boldsymbol{\alpha}_t^\top \mathbf{b}_\text{s} \) in~\eqref{Iterm} is neglected in comparison to \( \boldsymbol{\alpha}_t^\top \mathrm{diag}\{\mathbf{b}_\text{s}\} \boldsymbol{\alpha}_t \) for several reasons. First, this simplification improves tractability. Second, both terms attain their minimum when \( \boldsymbol{\alpha} = \mathbf{b}_\text{w} \), meaning that minimizing the quadratic form \( \boldsymbol{\alpha}^\top \mathrm{diag}\{\mathbf{b}_\text{s}\} \boldsymbol{\alpha} \) inherently suppresses the linear term \( \boldsymbol{\alpha}^\top \mathbf{b}_\text{s} \). Moreover, the quadratic term aligns with the widespread use of \( \ell_2 \)-norm-based objectives in optimization, offering smoother behavior compared to the \( \ell_1 \)-norm.

\textbf{Limitations:}
\begin{itemize}
\item {In the single-antenna setting, it requires intermediate synchronization; otherwise, the effective channel signs and the optimized aggregation weights in~\eqref{weightderror} may no longer match the actual received superposition.}
	\item The potential risk of aggregation bias and fairness degradation increases as the norm of the selected weight vector deviates from that of the equal-weight case.
	\item When aggregation weights are strictly dictated by the learning process, especially under high heterogeneity, such an approach may fail due to the lack of flexibility needed for effective adaptation.
\end{itemize}
\vspace{-5pt}
\section{{Digital and Hybrid Implementations of AirFL}}
\label{sec:digital_aircomp}

{The preceding sections classified AirFL according to three design principles: transmitter-side compensation, receiver-side equalization, and learning-aware aggregation weighting. Digital AirFL should be viewed through the same taxonomy, not as a fourth class. It realizes these principles through finite constellations, digital modulation, source quantization, channel coding, and packet-compatible interfaces. In particular, finite constellations can be interpreted as a quantized form of analog AirFL: devices transmit finite-resolution symbols whose wireless superposition still carries information about the desired aggregate. Under suitable quantization and labeling, the aggregation operation can also partially suppress independent quantization noise, in the same way that averaging reduces zero-mean perturbations across devices. At the same time, the use of discrete constellations enables reliability mechanisms that are not available in purely analog aggregation, such as constellation design, coding, and aggregation-oriented decoding. Consequently, digital AirFL changes the aggregation-error model: continuous analog distortion is partly replaced by quantization error, constellation-decision error, coding gain, and decoding-failure probability.}

{\textit{Local-CSIT digital AirFL.}
	One-bit broadband digital aggregation (OBDA)~\cite{gunduz_bpsk} follows the CSIT-aware principle. Devices quantize each gradient coordinate to one bit and transmit the resulting signs using BPSK, or equivalently 4-QAM through two orthogonal BPSK streams, over an OFDM interface. The server computes an over-the-air majority vote, so the recovered object is a sign-based FL update rather than a full-precision FedAvg average. Although OBDA is digital, it is not blind: it uses truncated channel inversion over OFDM subchannels to align the received signs. Therefore, OBDA inherits the local-CSIT AirFL features of transmitter-side compensation, local-CSI dependence, truncation loss, and sensitivity to CSI errors. Its value is compatibility with simple digital broadband radios and strong communication compression; its cost is the loss of amplitude information and its restriction to sign-based aggregation.}

{\textit{Fully blind digital AirFL.}
	The q-QAM blind FL scheme in~\cite{saeed}, called ChannelCompFed, follows the fully blind AirFL principle. Devices have no CSIT and do not perform channel inversion. Instead, they quantize local gradients and map them to q-QAM symbols, while a multi-antenna server uses receiver-side processing to mitigate fading and recover the mean. The key coding issue is that ordinary QAM labeling, such as Gray coding, is not computation-friendly: different tuples of transmitted symbols may superpose to the same received point while corresponding to different function values. ChannelCompFed therefore uses the ChannelComp principle of non-overlapping computation-oriented constellations~\cite{dig_aircomp}, with a closed-form SumComp-type q-QAM construction specialized to sum computation~\cite{sumcomp}. In this construction, the constellation labeling is chosen so that valid superposed constellation points correspond to valid aggregate values, and the decoder maps the received point to the nearest computation point. Thus, compared with OBDA, it supports higher-resolution aggregation beyond one-bit majority voting; compared with analog blind AirFL, it turns part of the analog distortion into a quantization--reliability tradeoff controlled by constellation design, coding, and receive diversity.}

{\textit{Weighted and extraction-based digital AirFL.}
	A broader digital viewpoint is provided by out-of-air computation (AirCPU)~\cite{aircpu}. In contrast to conventional AirComp, which pre-embeds a desired function into transmitted waveforms or relies on asymptotic averaging, AirCPU adopts an extraction principle: useful function representations are decoded from the received wireless superposition. Structured codes play a central role in this digital extraction view. In particular, AirCPU uses multi-layer nested lattice coding to progressively extract integer-coefficient function representations that can be combined or reused as side information.}

{Compute-Update FL (FedCPU)~\cite{my_dig} can be interpreted as an FL-specific single-layer lattice-coded realization of this extraction view and, at the same time, as a digital implementation of weighted AirFL. Model updates are normalized, dithered, lattice-quantized, and transmitted without CSIT or channel inversion. The whole process is end-to-end through joint source-channel coding~\cite{my_my}. The server decodes an integer combination of the transmitted lattice points, and this decoded combination is directly converted into the FL aggregate. Hence, unlike classical compute-and-forward~\cite{compute_forward}, the integer coefficients are not merely auxiliary communication objects for recovering individual messages; they define the aggregation weights. This gives FedCPU a coding gain through the lattice decoding region: moderate noise and interference do not necessarily perturb the aggregate unless they move the received point outside the correct decoding region. The resulting tradeoff is between quantization resolution, decoding reliability, and the learning quality of the induced integer weights.}

{\textit{Hybrid implementation.}
	Hybrid implementation should be interpreted as a deployment mechanism rather than a separate AirFL principle. One relevant line is hybrid analog--digital beamforming for AirComp, where analog RF combining and digital baseband processing reduce the number of RF chains in massive-MIMO receivers~\cite{zhai_hybrid_aircomp,zhai_twotime_hybrid_aircomp}. This supports the receiver-side processing needed by blind or weighted AirFL, but it is not a complete AirFL protocol by itself. Another relevant line is system-level analog--digital scheduling, where over-the-air aggregation is used for suitable devices or update components, while conventional digital links support control signaling, scheduling, synchronization, retransmission, or fallback transmission~\cite{adfl}. From a protocol perspective, an AirFL-compatible uplink can reuse synchronization and waveform structures already present in existing systems, but the scheduler must configure the aggregation resource differently: instead of assigning orthogonal resources for individual-message recovery, it allows the devices contributing to the same aggregate to transmit over a common subchannel or resource block in a controlled non-orthogonal manner. In this sense, hybrid implementation explains how AirFL can coexist with digital packet-based wireless systems without changing the three underlying AirFL design principles.}

\vspace{-5pt}
\section{{Comparative Evaluation}}
\label{sec:experimental_comparison}

\begin{figure}[t]
	\centering
	
	\begin{subfigure}[t]{0.46\linewidth}
		\centering
		\includegraphics[width=\linewidth]{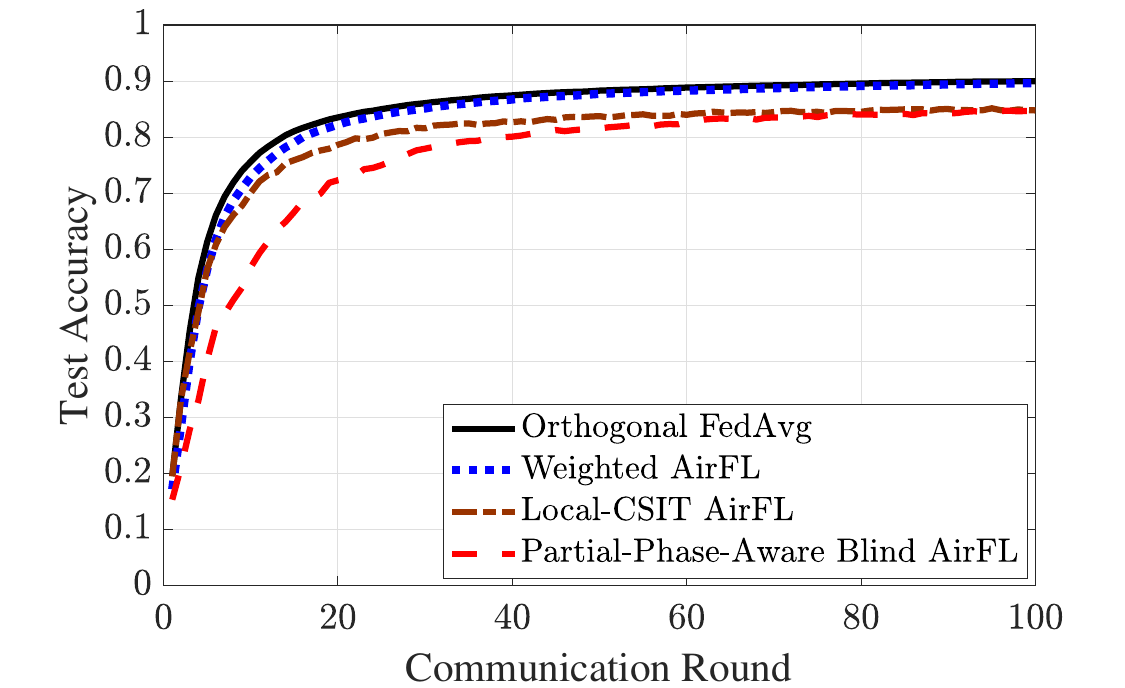}
		\caption{{MNIST test accuracy.}}
		\label{fig:mnist_test_accuracy}
	\end{subfigure}
	\hfill
	\begin{subfigure}[t]{0.46\linewidth}
		\centering
		\includegraphics[width=\linewidth]{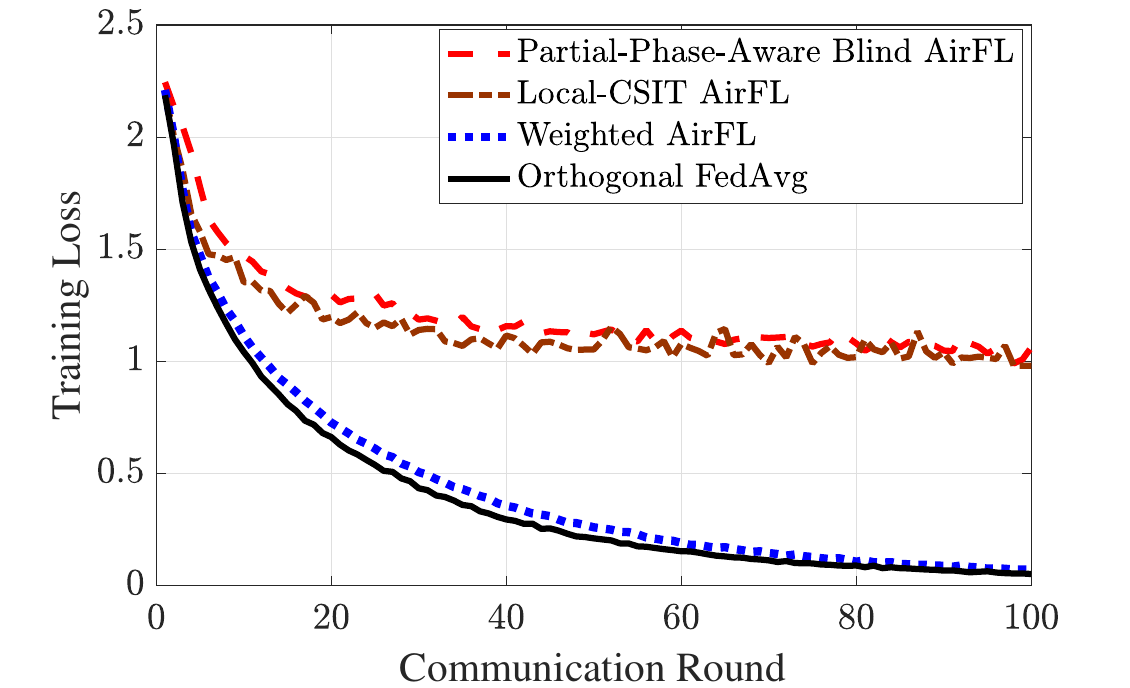}
		\caption{{CIFAR-10 training loss.}}
		\label{fig:cifar_training_loss}
	\end{subfigure}
	\vspace{-10pt}
	\caption{{Comparison of single-antenna AirFL schemes on MNIST and CIFAR-10.}}
	\label{fig:airfl_mnist_cifar}
	\vspace{-10pt}
\end{figure}

\begin{figure}[t]
	\centering
	\includegraphics[width=0.46\linewidth]{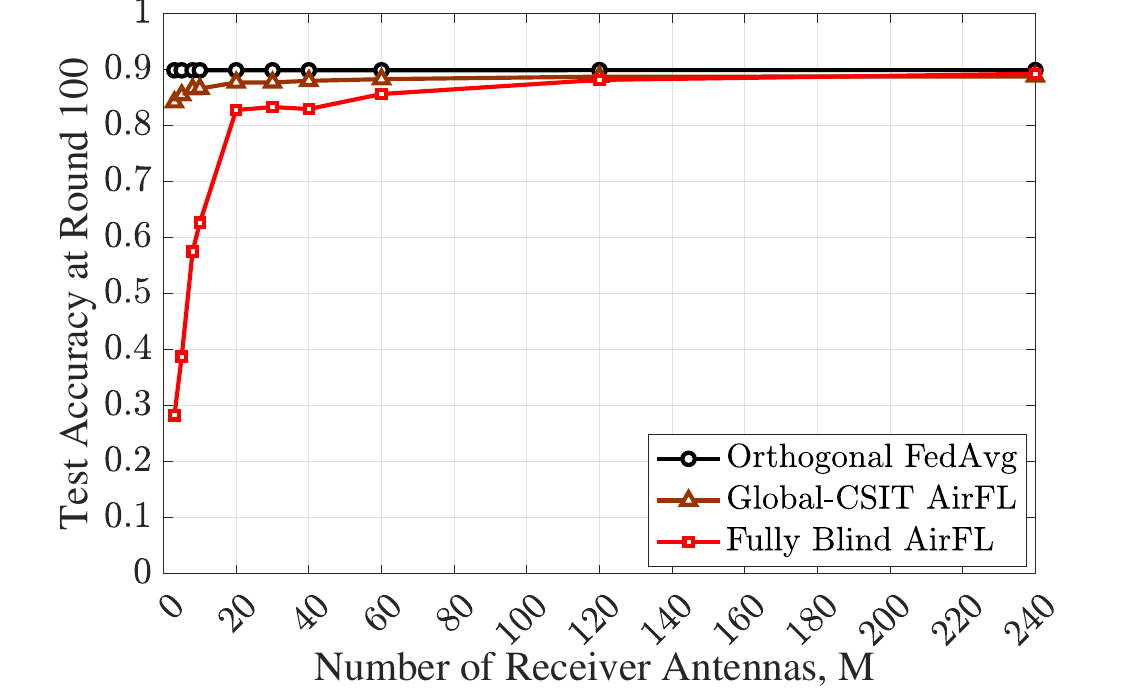}
	\vspace{-8pt}
	\caption{{Final MNIST test accuracy of multiple-antenna AirFL schemes versus number of antennas.}}
	\label{fig:accuracy_M}
	\vspace{-20pt}
\end{figure}

{We complement the analytical discussion with numerical comparisons organized by the server architecture required by the representative AirFL mechanisms. The purpose is not to establish a universal ranking, since the schemes rely on different assumptions on CSI, synchronization, antenna resources, and aggregation flexibility. Instead, the goal is to illustrate how the main aggregation mechanisms behave under controlled learning and wireless settings. Fig.~\ref{fig:airfl_mnist_cifar} compares the single-server-antenna AirFL schemes considered in this tutorial, namely Local-CSIT AirFL, Partial-Phase-Aware Blind AirFL, and WAFeL. Orthogonal FedAvg is also shown as an error-free digital benchmark, where local updates are delivered over orthogonal resources and averaged exactly at the server, requiring more resources than AirFL in proportion to the number of devices. Fig.~\ref{fig:accuracy_M} compares the multiple-antenna-server AirFL schemes, namely Global-CSIT AirFL and Fully Blind AirFL, with Orthogonal FedAvg as the reference.}

{For Fig.~\ref{fig:mnist_test_accuracy}, we use $K=30$ devices, each with $200$ disjoint i.i.d. MNIST training samples. The mini-batch size is $20$, and each device performs $20$ local SGD steps per communication round. The optimizer is SGD with learning rate $0.01$ and no momentum. The result is averaged over $10$ Monte Carlo runs. The learning model is an intentionally lightweight MLP consisting of an input flattening layer, one fully connected hidden layer with $64$ ReLU neurons, and a softmax output layer. This lightweight design focuses the comparison on the relative behavior of the wireless aggregation mechanisms rather than on achieving state-of-the-art MNIST accuracy.}

{For Fig.~\ref{fig:cifar_training_loss}, we keep $K=30$ but use CIFAR-10 with $1600$ disjoint i.i.d. training samples per device. The CNN has two $32$-filter convolutional layers with $3\times3$ kernels, same padding, and ReLU activation, followed by $2\times2$ max-pooling and dropout with factor $0.25$, two $64$-filter convolutional layers with $3\times3$ kernels, same padding, and ReLU activation, followed by $2\times2$ max-pooling and dropout with factor $0.25$, a flattening layer, a dense layer with $256$ ReLU neurons, dropout with factor $0.5$, and a softmax output layer. The mini-batch size is $32$, and each device performs $150$ local SGD steps per communication round. The optimizer is SGD with learning rate $0.01$ and momentum $0.9$. One Monte Carlo run is reported to make the per-realization variability of the schemes visible.}

{The wireless parameters in Fig.~\ref{fig:airfl_mnist_cifar} are common to both datasets: AWGN variance $\sigma_z^2=1$, transmit power $P=10$, and SNR $=10$ dB. For Local-CSIT AirFL, the channel-inversion threshold is $\theta=0.2$. For WAFeL, the MSE threshold is selected between the minimum-MSE and equal-weight points using target factor $0.5$.}

{Fig.~\ref{fig:mnist_test_accuracy} shows that WAFeL almost overlaps with Orthogonal FedAvg and reaches nearly the same final test accuracy. This indicates that, for the lightweight MNIST model, adaptive aggregation weights and receiver equalization can make analog aggregation behave close to exact digital averaging. Local-CSIT AirFL also learns, but remains below WAFeL because threshold-based channel inversion averages only over the devices that pass the channel threshold in each round. Partial-Phase-Aware Blind AirFL has the weakest MNIST performance because its aggregate remains directly shaped by random effective channel coefficients. Nevertheless, its nonzero learning progress shows that useful training is still possible even with very limited channel knowledge and minimal aggregation correction.}

{Fig.~\ref{fig:cifar_training_loss} shows a clearer separation among the same single-server-antenna AirFL schemes. Orthogonal FedAvg and WAFeL reduce the training loss steadily, with WAFeL remaining close to the digital reference. In contrast, Local-CSIT AirFL and Partial-Phase-Aware Blind AirFL show much larger losses and stronger round-to-round variability. For Local-CSIT AirFL, this variability comes from the random active set induced by channel truncation. For Partial-Phase-Aware Blind AirFL, it comes from uncontrolled channel-dependent weighting of the aggregate. The effect is much more visible on CIFAR-10 because the deeper CNN and momentum-based local training make the optimization trajectory more sensitive to aggregation distortion.}

{Thus, Fig.~\ref{fig:airfl_mnist_cifar} shows that the same wireless aggregation mechanisms can behave differently depending on the learning task and model. In the lightweight MNIST setting, all schemes can still learn useful models. In the CIFAR-10 setting, random device truncation and uncontrolled channel-weighted aggregation produce a much larger optimization penalty, while WAFeL remains comparatively stable by adapting the aggregation weights to the instantaneous wireless aggregation condition. This should not be interpreted as proving that CSIT-aware AirFL is intrinsically inferior; rather, it shows that, in this representative single-server-antenna comparison, aggregation-weight adaptation provides an effective way to trade exact averaging for wireless robustness. Similar weighting ideas could also be incorporated into CSIT-aware designs, at the cost of additional coordination, feedback, and fairness management.}

{Fig.~\ref{fig:accuracy_M} uses the same MNIST setup as Fig.~\ref{fig:mnist_test_accuracy}. The difference is that the server is equipped with multiple antennas, and the compared AirFL schemes are Global-CSIT AirFL and Fully Blind AirFL. The reported metric is the MNIST test accuracy at communication round 100 for different numbers of server antennas. Global-CSIT AirFL uses matching-pursuit scheduling with equal scheduling weights, MSE tolerance $0.10$, and subset-cutting parameter $0.05$~\cite{bereyhi}.}

{Fig.~\ref{fig:accuracy_M} shows that Global-CSIT AirFL stays close to Orthogonal FedAvg even with a small number of receive antennas, while Fully Blind AirFL starts with low accuracy and improves rapidly as $M$ increases. This highlights the different role of multiple antennas in the two schemes: Global-CSIT AirFL uses centralized CSI-dependent scheduling and receive processing to shape the aggregate in a potentially optimal manner, whereas Fully Blind AirFL relies on statistical spatial averaging to reduce residual aggregation error. Therefore, the practical question for Fully Blind AirFL is not whether infinitely many antennas are needed for exact averaging, but how large a finite receive array must be for the residual aggregation error to become acceptable for learning. In this setup, achieving a target test accuracy of $80\%$ requires at least 20 antennas.}

{The simulation code is publicly available online.\footnote{{\url{https://github.com/seyaa90/airfl-spm-simulation-code}}}} A comparative overview is also provided in Table~\ref{tab:comparison_revised}, drawing on the above discussions in the preceding sections and the experimental observations above.

\begin{table}[!htbp]
	\centering
	\fontsize{6.7pt}{7.3pt}\selectfont
	\setlength{\tabcolsep}{1.3pt}
	\renewcommand{\arraystretch}{1.195}
	\vspace{-10pt}
	\caption{{Comparison of representative AirFL approaches.}}
	\vspace{-7pt}
	\label{tab:comparison_revised}
	\begin{tabularx}{0.98\textwidth}{|p{1.75cm}|Y|p{1.15cm}|Y|Y|p{2.35cm}|}
		\hline
		\textbf{Approach} 
		& \textbf{Key Design Element} 
		& \textbf{Sync. Level} 
		& \textbf{CSI / Channel Knowledge} 
		& \textbf{Main Limitation} 
		& \textbf{Main Design Complexity} \\
		\hline
		
		Orthogonal FedAvg benchmark
		& Separate packet-based transmissions followed by digital averaging
		& Packet-level
		& CSIR for individual-link decoding
		& Requires $K$ orthogonal communication resources and does not exploit waveform superposition
		& No AirFL optimization; standard decoding and server averaging \\
		\hline
		
		Local-CSIT AirFL
		& Device-side channel inversion, threshold-based power control, and hard device selection
		& Fine
		& Local effective CSIT at each device
		& Sensitive to channel aging and synchronization errors; weak-channel truncation limits participation and may harm fairness; representative scalar design is tied to single-antenna aggregation
		& $\mathcal{O}(1)$ per device; $\mathcal{O}(K)$ total thresholding/precoding \\
		\hline
		
		Global-CSIT AirFL
		& Centralized device selection, receive equalization, denormalizing, and device-side power control
		& Fine
		& Global effective CSI at the server; device-specific precoder feedback to transmitters
		& CSI acquisition, feedback, channel-aging, synchronization, power-control, and optimization overhead; hard selection may affect fairness
		& DC: $\mathcal{O}\!\big(K((M^{2}+K)^{3}+M^{6})\big)$; MP: $\mathcal{O}(K^{2}M^{2})$ \\
		\hline
		
		Fully blind AirFL
		& Constant-power transmission with massive-MIMO receiver equalization and statistical averaging
		& Coarse
		& No CSIT; receiver uses CSIR and/or channel statistics for aggregation
		& Requires sufficiently large antenna arrays and restrictive i.i.d. channel assumptions; finite $M$ leaves residual interference/noise and possible aggregation bias
		& No iterative optimization; linear receiver combining over antennas \\
		\hline
		
		Partial-phase-aware blind AirFL
		& Constant-power transmission with quadrant phase compensation and imposed channel-weighted aggregation
		& Intermediate
		& Limited phase information at devices; no full channel inversion
		& Aggregation weights are induced by channel gains, leading to biased aggregation and fairness issues; convergence analysis relies on strong channel, interference, and learning assumptions
		& No iterative optimization; simple phase compensation \\
		\hline
		
		Weighted AirFL
		& Learning-aware aggregation-weight design with receiver equalization
		& Intermediate in the single-antenna phase-compensated realization
		& CSIR/effective channel matrix at the server; limited phase information at devices in the single-antenna realization
		& Unequal optimized weights create a bias--fairness tradeoff; flexibility may be limited when the learning task requires prescribed aggregation weights
		& Weight selection: $\mathcal{O}(K^{3})$; low-dimensional equalization after weights are chosen \\
		\hline
	\end{tabularx}
\end{table}

\vspace{-25pt}
\section{Conclusions and Future Directions}

{
	This article presented AirFL as a signal-processing problem at the interface between wireless computation and distributed learning. Rather than viewing the wireless channel only as a communication impairment, AirFL treats waveform superposition as a computational resource for model aggregation. The key design question is therefore not only how accurately a receiver can estimate a transmitted signal, but how the received superposition can be shaped into a learning-useful aggregate. This perspective leads to three complementary design principles: compensating the channel at the transmitters, extracting the aggregate at the receiver, and adapting the learning aggregation rule itself.
	
	A central lesson is that AirFL should not be evaluated only through communication metrics such as aggregation MSE. Since the received aggregate directly drives the learning dynamics, signal-processing choices must ultimately be assessed through convergence, fairness, device participation, and deployment feasibility. This creates a new design space in which channel knowledge, synchronization, receiver processing, aggregation weights, coding, and learning objectives are coupled. The following directions summarize open challenges for future AirFL research beyond the individual schemes discussed above.
	
	\begin{itemize}
		
	\item \textit{Synchronization-aware aggregation design:}
	Future AirFL systems need synchronization models that are directly tied to learning performance. Residual timing offsets, carrier-frequency offsets, sampling-frequency offsets, phase noise, and hardware calibration errors should not be treated only as communication impairments; they should be mapped to aggregation bias, distortion, and convergence degradation. A key challenge is to design aggregation estimators, equalizers, and weights that remain reliable when effective channels are partially known, outdated, or varying across model coordinates. This calls for joint designs of pilot signaling, timing advance, guard intervals, phase tracking, receiver-side misalignment compensation, robust equalization, and adaptive aggregation weights. Such synchronization-aware designs are especially important for mobile devices, wideband channels, low-cost oscillators, and long model-update transmissions.
		
		\item \textit{Local-CSIT AirFL with multiple-antenna edge devices and servers:}
		Extending local-CSIT AirFL beyond scalar channel inversion becomes more challenging in multi-antenna settings. Unlike scalar inversion, multi-dimensional channel inversion is not uniquely defined for aggregation because it depends on the chosen transmit and receive spatial directions. With multiple-antenna edge devices, each transmitter observes only its own local channel vector or matrix and must choose a transmit beamformer without knowing how the other devices' beams will combine at the receiver. At the same time, the server must design a receive equalizer that extracts a common aggregate from the superposed multi-antenna signals. This creates several technical hurdles: decentralized transmit beamforming, aggregation-subspace alignment, residual inter-device interference suppression, per-antenna power constraints, limited RF chains, hardware calibration, phase coherence across antennas and devices, heterogeneous antenna configurations, and channel aging. Future work should develop local-CSI beamforming and robust receive-processing methods that preserve the AirFL learning objective without requiring global CSI.
		
		\item \textit{Blind AirFL with finite and imperfect arrays:}
		Receiver-side processing can reduce the burden on edge devices, but practical receivers have finite antenna arrays, spatially correlated channels, unequal path losses, low-resolution converters, nonlinear hardware, and imperfect calibration. Future work should therefore move beyond ideal massive-MIMO averaging and design finite-dimensional equalizers that are robust to non-i.i.d. fading and hardware impairments. Important directions include covariance-aware combining, regularized equalization, learning-aware receiver design, and antenna-scaling laws that quantify how many spatial degrees of freedom are needed to achieve a target learning performance under realistic propagation conditions.
		
\item \textit{Adaptive weighted aggregation beyond fixed averaging:}
Future AirFL research should move from recovering a fixed average to designing aggregation rules that adapt jointly to wireless conditions and learning needs. One direction is weighted CSIT-aware AirFL, where device selection, transmit precoding, receive equalization, and aggregation weights are co-designed under channel uncertainty, feedback limits, power constraints, and fairness requirements. More generally, adaptive aggregation weights can be incorporated into different AirFL classes and jointly optimized with their native design variables, increasing the available degrees of freedom for improving learning performance, robustness, and fairness. Another direction is long-horizon aggregation control, where weights account not only for instantaneous channel quality and aggregation distortion, but also for device participation history, energy state, persistent channel disadvantage, data diversity, and learning usefulness across rounds. This raises concrete questions on how to quantify beneficial aggregation bias, how to protect statistically important but channel-disadvantaged devices, how to balance short-term communication reliability with long-term convergence and fairness, and how to keep adaptive weights stable under channel aging and non-i.i.d. data.
		
\item \textit{Digital, hybrid, and generalized computation for AirFL:}
A key future direction is to move AirFL from idealized analog aggregation toward wireless computation mechanisms compatible with practical digital and hybrid protocol stacks. This requires quantization, modulation, coding, scheduling, retransmission, control signaling, fallback links, and adaptive AirFL/orthogonal switching to be optimized for the final learning update rather than only individual-message recovery. Open problems include learning-aware quantization, computation-oriented channel coding, joint source--channel coding for model-update transmission and aggregation, hybrid analog--digital operation, and reliability analysis that maps quantization, detection, decoding, and delay errors to aggregation distortion, update bias, and convergence degradation. Beyond digitizing existing AirFL schemes, out-of-air computation~\cite{aircpu} enables generalized modes where AirFL is not limited to one-shot averaging but can support collective and successive computation over wireless superposition. Multiple extracted function representations may then be combined, selected, or reused across antennas, receivers, clusters, or rounds to form learning-useful aggregates.

\item \textit{Robustness, security, and privacy in waveform-domain learning:}
The superposition property that enables AirFL also creates new vulnerabilities: malicious devices may manipulate the aggregate without being individually decoded, interference may corrupt the update, and analog transmissions may leak information about local models. Future research should therefore go beyond simple linear aggregation, such as sums or averages, and study robust nonlinear rules, including geometric median, trimmed mean, and coordinate-wise median. A key challenge is how to realize or approximate such robust functions over the air, since the wireless channel naturally provides only linear superposition. Promising directions include iterative over-the-air protocols, analog--digital robust aggregation, sketching/projection methods, nonlinear receiver processing, physical-layer privacy, analog masking, secure beamforming, and anomaly detection, all jointly evaluated with aggregation distortion and learning convergence.

\item \textit{Personalized, clustered, and heterogeneity-robust AirFL:}
Conventional AirFL mainly computes one global over-the-air aggregate, which may be inadequate under strong statistical, computational, task, and channel heterogeneity. Building on initial studies of over-the-air clustered FL and personalized AirFL~\cite{sami_clustered,li_personalized}, future work should connect personalization and clustering with the underlying AirFL mechanism rather than treating them only as learning-layer extensions. In CSIT-aware AirFL, this may require channel-dependent grouping, beamforming, power control, and cluster selection. In blind AirFL, receiver-side separation, spatial processing, or orthogonalized aggregation resources may be needed to avoid undesired mixing of different clusters or tasks. In weighted AirFL, adaptive aggregation weights can reflect both learning similarity and wireless reliability. Unlike conventional FL, heterogeneity in AirFL affects not only the learning objective but also the wireless aggregation itself: devices with similar tasks may have incompatible channels, while devices with favorable channels may have dissimilar data or objectives. Future designs should therefore jointly optimize device grouping, aggregation-slot sharing, cluster/task separation through time, frequency, code, beam, or receiver processing, and aggregation weights that balance statistical similarity, computation reliability, fairness, and channel-induced distortion. This can move AirFL from one global wireless average toward structured over-the-air aggregates that preserve shared knowledge while enabling local specialization across heterogeneous devices.

\item \textit{Architectures beyond a single shared aggregation interface:}
Direct AirFL is best suited to devices sharing a common wireless aggregation interface, while many edge-learning systems are multi-cell, hierarchical, mobile, or geographically distributed. Future work should integrate AirFL with hierarchical FL, cell-free and distributed MIMO, coordinated multi-point reception, RIS-assisted aggregation, UAV/non-terrestrial platforms, and multi-server edge coordination. Key challenges include receiver cooperation, fronthaul-efficient aggregation, inter-cluster interference control, mobility-aware scheduling, and graph-dependent convergence. Decentralized device-to-device AirFL, based on peer-to-peer superposition, consensus, or gossip protocols, is another promising direction.
		
	\end{itemize}
	
Overall, AirFL points toward programmable wireless computation for distributed AI: a framework in which wireless superposition is not merely tolerated, but deliberately shaped into learning-useful computation under practical constraints.
}

\vfill


\vspace{-10pt}
\begin{thebibliography}{1}
	\bibliographystyle{IEEEtran}
	\bibitem{mcmahan}
	B. McMahan, E. Moore, D. Ramage, S. Hampson, and B. A. Arcas, "Communication-efficient learning of deep networks from decentralized data," in \emph{Proc. AISTATS}, pp. 1273--1282, 2017.
	
	\bibitem{nazer}
	B. Nazer and M. Gastpar, "Computation over multiple-access channels,"
	\emph{IEEE Trans. Inf. Theory}, vol. 53, no. 10, pp. 3498--3516, Oct. 2007.
	
	\bibitem{gold}
	M. Goldenbaum, H. Boche, and S. Stanczak, "Harnessing interference for analog function computation in wireless sensor
	networks," \emph{IEEE Trans Signal Proc.}, vol. 61, no. 20, pp. 4893--4906, 2013.
	
	\bibitem{netmag}
	S. M. Azimi-Abarghouyi, M. Bennis, and L. Tassiulas, "Hierarchical federated learning for networked AI: From communication saving to architecture-aware design," available on arXiv: 2605.00931 
	
	\bibitem{localCSIT}
	G. Zhu, Y. Wang, and K. Huang, "Broadband analog aggregation for low-latency federated edge learning," \emph{IEEE Trans. Wireless Commun.}, vol. 19, no. 1, pp. 491--506, Jan. 2020.
	
	\bibitem{globalCSIT}
	K. Yang, T. Jiang, Y. Shi, and Z. Ding, "Federated learning via over-the-air computation," \emph{IEEE Trans. Wireless Commun.}, vol. 19, no. 3, pp. 2022--2035, Mar. 2020.
	
	\bibitem{blind}
	M. Mohammadi Amiri, T. M. Duman, D. Gunduz, S. R. Kulkarni, and H. V. Poor, "Blind federated edge learning," \emph{IEEE Trans. Wireless Commun.}, vol. 20, no. 8, pp. 5129--5143, Aug. 2021.
	
	\bibitem{wafel}
	S. M. Azimi-Abarghouyi and L. Tassiulas, "Over-the-air federated learning via weighted aggregation," \emph{IEEE Trans. Wireless Commun.},  vol. 23, no. 12, pp. 18240--18253, Dec. 2024.
	
	\bibitem{smith}
	T. Li, A. K. Sahu, A. Talwalkar, V. Smith, "Federated learning: Challenges, methods, and future directions," \emph{IEEE Signal Process. Mag.}, vol. 37, no. 3, pp. 50--60, May 2020.
	
	\bibitem{israel}
	T. Gafni, N. Shlezinger, K. Cohen, Y. C. Eldar, and H. V. Poor, "Federated learning: A signal processing perspective," \emph{IEEE Signal Process. Mag.}, vol. 39, no. 3, pp. 14--41, May 2022.
	
	\bibitem{alphan}
	A. Sahin and R. Yang "A survey on over-the-air computation," \emph{IEEE Commun. Surv. Tutor.}, vol. 25, no. 3, pp. 1877--1908, 2023.
	
	\bibitem{saeedmag}
	A. Perez-Neira, M. Martinez-Gost, A. Sahin,
	S. Razavikia, C. Fischione, and K. Huang, "Waveforms for computing over the air," \emph{IEEE Signal Process. Mag.}, vol. 42, no. 2, pp. 57--77, Mar. 2025.
	
	\bibitem{over1}
	Z. Chen, H. H. Yang, and T. Q. S. Quek, "Edge intelligence over the air: Two faces
	of interference in federated learning,"  \emph{IEEE Commun. Mag.}, vol. 61, no. 12, pp. 62--68, Dec. 2023.
	
	\bibitem{over2}
	X. Cao, Z. Lyu, G. Zhu, J. Xu, L. Xu, and S. Cui, "An overview on over-the-air
	federated edge learning," \emph{IEEE Wirel. Commun.}, vol. 31, no. 3, June 2024.
	
	\bibitem{over3}
	M. Chen, D. Gunduz, K. Huang, W. Saad, M. Bennis, A. V. Feljan, and H. V. Poor, "Distributed learning in wireless networks: Recent progress and future challenges," \emph{IEEE J. Sel. Areas Commun.}, pp. 1--26, 2021.
	
	\bibitem{over4}
	H. Hellstrom, J. M. Barros da Silva Jr., M. M. Amiri, M. Chen, V. Fodor,
	H. V. Poor, and C. Fischione, "Wireless for machine learning: A
	survey," \emph{Foundations and Trends in Signal Processing}, vol. 15, no. 4,
	pp. 290--399, 2022.
	
	\bibitem{tao_feel6g}
	{M. Tao, Y. Zhou, Y. Shi, J. Lu, S. Cui, J. Lu, and K. B. Letaief,
		"Federated edge learning for 6G: Foundations, methodologies, and applications,"
		\emph{Proc. IEEE}, vol. 113, no. 9, pp. 1075--1113, Sept. 2025.}
	
	\bibitem{khabin}
	X. Cao, G. Zhu, J. Xu, K. Huang, "Optimized power control for over-the-air computation in fading channels," \emph{IEEE Trans. Wireless Commun.}, vol. 19, no. 11, pp. 7498--7513, Nov. 2020.
	
	\bibitem{khabin2}
	X. Cao, G. Zhu, J. Xu, and K. Huang, "Cooperative interference
	management for over-the-air computation networks," \emph{IEEE Trans.
		Wireless Commun.}, vol. 20, no. 4, pp. 2634--2651, Apr. 2020.
	
	\bibitem{stich_local_sgd}
	{S. U. Stich, "Local SGD converges fast and communicates little," in \emph{Proc. Int. Conf. Learn. Represent. (ICLR)}, 2019.}
	
	\bibitem{mahmood_time_sync}
	{A. Mahmood, M. I. Ashraf, M. Gidlund, J. Torsner, and J. Sachs,
		"Time synchronization in 5G wireless edge: Requirements and solutions for critical-MTC,"
		\emph{IEEE Commun. Mag.}, vol. 57, no. 12, pp. 45--51, Dec. 2019.}
	
	\bibitem{guo_waveform}
	{H. Guo, Y. Zhu, H. Ma, V. K. N. Lau, K. Huang, X. Li, H. Nong, and M. Zhou,
		"Over-the-air aggregation for federated learning: Waveform superposition and prototype validation,"
		\emph{J. Commun. Inf. Netw.}, vol. 6, no. 4, pp. 429--442, Dec. 2021.}
	
	\bibitem{airshare}
	{O. Abari, H. Rahul, D. Katabi, and M. Pant,
		"AirShare: Distributed coherent transmission made seamless,"
		in \emph{Proc. IEEE INFOCOM}, Hong Kong, China, Apr. 2015.}
	
	\bibitem{shao_misaligned}
	{Y. Shao, D. G\"und\"uz, and S. C. Liew,
		"Federated edge learning with misaligned over-the-air computation,"
		\emph{IEEE Trans. Wireless Commun.}, vol. 21, no. 6, pp. 3951--3964, Jun. 2022.}
	
	\bibitem{shao_bayesian}
	{Y. Shao, D. G\"und\"uz, and S. C. Liew,
		"Bayesian over-the-air computation,"
		\emph{IEEE J. Sel. Areas Commun.}, vol. 41, no. 3, pp. 589--606, Mar. 2023.}
	
	\bibitem{gunduz_bpsk}
	G. Zhu, Y. Du, D. Gunduz, and K. Huang, "One-bit over-the-air aggregation for communication-efficient federated edge learning: Design and convergence analysis", \emph{IEEE Trans. Wireless Commun.}, vol. 20, no. 3, pp. 2120--2135, Mar. 2021. 
	
	\bibitem{sahin_dfts_ofdm}
	{A. Sahin, B. Everette, and S. S. M. Hoque,
		"Over-the-air computation with DFT-spread OFDM for federated edge learning,"
		in \emph{Proc. IEEE Wireless Commun. Netw. Conf. (WCNC)}, Austin, TX, USA, Apr. 2022.}
	
	\bibitem{zhao_digital_async}
	{X. Zhao, L. You, R. Cao, Y. Shao, and L. Fu,
		"Broadband digital over-the-air computation for asynchronous federated edge learning,"
		in \emph{Proc. IEEE Int. Conf. Commun. (ICC)}, Seoul, South Korea, May 2022.}
	
	\bibitem{you_digital_tmc}
	{L. You, X. Zhao, R. Cao, Y. Shao, and L. Fu,
		"Broadband digital over-the-air computation for wireless federated edge learning,"
		\emph{IEEE Trans. Mobile Comput.}, vol. 23, no. 5, pp. 5491--5505, May 2024.}
	
	\bibitem{khabin_sg}
	Z. Lin, X. Li, V. K. Lau, Y. Gong, and K. Huang, "Deploying federated learning in large-scale cellular
	networks: Spatial convergence analysis," \emph{IEEE Trans. Wireless Commun.}, vol. 21, no. 3, pp. 1542--1556, Mar. 2022.
	
	\bibitem{khabin_noise}
	Z. Zhang, G. Zhu, R. Wang, V. K. Lau, and K. Huang, "Turning channel noise into an accelerator for over-the-air principal component analysis," \emph{IEEE Trans. Wireless Commun.}, vol. 21, no. 10, pp. 7926--7941, Oct. 2021.
	
	\bibitem{azimi_hiersg}
	S. M. Azimi-Abarghouyi and V. Fodor, "Scalable hierarchical over-the-air federated learning," \emph{IEEE Trans. Wireless Commun.}, vol. 23, no. 8, pp. 8480--8496, Aug. 2024.
	
	\bibitem{sery_hetero}
	{T. Sery, N. Shlezinger, K. Cohen, and Y. C. Eldar, "Over-the-air federated learning from heterogeneous data," \emph{IEEE Trans. Signal Process.}, vol. 69, pp. 3796--3811, 2021.}
	
	\bibitem{bereyhi} 
	A. Bereyhi, A. Vagollari, S. Asaad, R. R. Muller, W. Gerstacker, and
	H. V. Poor, "Device scheduling in over-the-air federated learning via matching pursuit," \emph{IEEE Trans. Signal Process.}, vol. 71, pp. 2188--2203, June 2023. 
	
	\bibitem{latif} 
	Z. Wang, J. Qiu, Y. Zhou, Y. Shi, L. Fu, W. Chen, and K. B. Letaief,
	"Federated learning via intelligent reflecting surface," \emph{IEEE Trans. Wireless Commun.}, vol. 21, no. 2, pp. 808--822, Feb. 2022.
	
	\bibitem{ng}
	M. Kim, A. L. Swindlehurst, and D. Park, "Beamforming vector design and device selection
	in over-the-air federated learning," \emph{IEEE Trans. Wireless Commun.}, vol. 22, no. 11, pp. 7464--7477, Nov. 2023.
	
	\bibitem{saeed}
	S. Razavikia, J. M. Barros da Silva Jr., and C. Fischione, "Blind federated learning via over-the-air q-QAM," \emph{IEEE Trans. Wireless Commun.}, vol. 23, no. 12, pp. 19570--19586, Dec. 2024.
	
	\bibitem{turky_adc}
	B. Tegin and T. M. Duman, "Blind federated learning at the wireless edge with low-resolution ADC and DAC," \emph{IEEE Trans. Wireless Commun.}, vol. 20, no. 12, pp. 7786--7798, Dec. 2021.
	
	\bibitem{turky_time}
	B. Tegin and T. M. Duman, "Federated learning with over-the-air aggregation over time-varying channels," \emph{IEEE Trans. Wireless Commun.}, vol. 22, no. 8, pp. 5671--5684, Aug. 2023.
	
	\bibitem{hier_deniz}
	O. Aygun, M. Kazemi, D. Gunduz, T. M. Duman, "Over-the-air federated edge learning with hierarchical clustering," \emph{IEEE Trans. Wireless Commun.}, vol. 23, no. 12, pp. 17856--17871, Dec. 2024.
	
	\bibitem{blind_simple1}
	T. Sery and K. Cohen, "On analog gradient descent learning over
	multiple access fading channels," \emph{IEEE Trans. Signal Process.}, vol. 68, pp. 2897--2911, 2020.
	
	\bibitem{blind_simple2}
	R. Paul, Y. Friedman, and K. Cohen, "Accelerated gradient descent
	learning over multiple access fading channels," \emph{IEEE J. Sel. Areas
		Commun.}, vol. 40, no. 2, pp. 532--547, Feb. 2022.
	
	\bibitem{blind_interference}
	H. H. Yang, Z. Chen, T. Q. S. Quek, and H. V. Poor, "Revisiting analog
	over-the-air machine learning: The blessing and curse of interference,"
	\emph{IEEE J. Sel. Topics Signal Process.}, vol. 16, no. 3, pp. 406--419,
	Apr. 2022.
	
	\bibitem{tailtheory}
	L. Clavier, T. Pedersen, I. Rodriguez, M. Lauridsen, and M. Egan,
	"Experimental evidence for heavy tailed interference in the IoT," \emph{IEEE
		Commun. Lett.}, vol. 25, no. 3, pp. 692--695, Mar. 2021.
	
	\bibitem{tailexperiment}
	D. Middleton, "Statistical-physical models of electromagnetic interference," \emph{IEEE Trans. Electromagn. Compat.}, no. 3, pp. 106-127, Aug.
	1977.
	
	\bibitem{wafel_conf}
	S. M. Azimi-Abarghouyi, L. Tassiulas, and C. Fischione, "Weighted over-the-air federated learning," \emph{IEEE ICMLCN}, Barcelona, Spain, May 2025.
	
	\bibitem{w_japan}
	K. Sato and K. Ishibashi, "Adaptively weighted averaging over-the-air
	computation and its application to
	distributed Gaussian process regression,"  \emph{IEEE Trans. Cogn. Commun. Netw.}, vol. 12, pp. 480--496, April. 2025.
	
	\bibitem{dig_aircomp}
	S. Razavikia, J. M. Barros da Silva Jr., and C. Fischione, "ChannelComp: A general method for computation by communications," \emph{IEEE Trans. Commun.}, vol. 72, no. 2, pp. 692--706, Feb. 2024.
	
	\bibitem{sumcomp}
	{S. Razavikia, J. M. B. da Silva Jr., and C. Fischione, "SumComp: Coding for digital over-the-air computation via the ring of integers," \emph{IEEE Trans. Commun.}, vol. 73, no. 2, pp. 752--767, Feb. 2025.}
	
	\bibitem{aircpu}
	{S. M. Azimi-Abarghouyi, "Out-of-air computation: Enabling structured extraction from wireless superposition," available on arXiv: 2604.04312}
	
	\bibitem{my_dig}
	S. M. Azimi-Abarghouyi and L. R. Varshney, "Compute-update federated learning: A lattice coding approach," \emph{IEEE Trans. Signal Process.},
	vol. 72, pp. 5213--5227, Nov. 2024. 
	
	\bibitem{my_my}
	S. M. Azimi-Abarghouyi and L. R. Varshney, "Federated learning via lattice joint source-channel coding," \emph{IEEE ISIT}, Athens, Greece, July 2024.
	
	
	\bibitem{compute_forward}
	{B. Nazer and M. Gastpar, "Compute-and-forward: Harnessing interference through structured codes," \emph{IEEE Trans. Inf. Theory}, vol. 57, no. 10, pp. 6463--6486, Oct. 2011.}
	
	\bibitem{zhai_hybrid_aircomp}
	{X. Zhai, X. Chen, J. Xu, and D. W. K. Ng, "Hybrid beamforming for massive MIMO over-the-air computation," \emph{IEEE Trans. Commun.}, vol. 69, no. 4, pp. 2737--2751, Apr. 2021.}
	
	\bibitem{zhai_twotime_hybrid_aircomp}
	{X. Zhai, X. Chen, and Y. Cai, "Power minimization for massive MIMO over-the-air computation with two-timescale hybrid beamforming," \emph{IEEE Wireless Commun. Lett.}, vol. 10, no. 4, pp. 873--877, Apr. 2021.}
	
	\bibitem{adfl}
	{M. F. Ul Abrar and N. Michelusi, "Analog-digital scheduling for federated learning: A communication-efficient approach," in \emph{Proc. 57th Asilomar Conf. Signals, Systems, and Computers (ACSSC)}, pp. 53--58, 2023.}
	
	\bibitem{sami_clustered}
	{H. U. Sami and B. Guler, "Over-the-air clustered federated learning," \emph{IEEE Trans. Wireless Commun.}, vol. 23, no. 7, pp. 7877--7893, Jul. 2024.}
	
	\bibitem{li_personalized}
	{Z. Li, Z. Chen, T. Q. S. Quek, and H. H. Yang, "Personalized federated learning over the air," \emph{IEEE Trans. Wireless Commun.}, vol. 24, no. 11, pp. 9509--9523, Nov. 2025.}
	
\end{thebibliography}
\end{document}